\documentclass[]{aastex631}
\graphicspath{{Figures/}}

\usepackage{bm}
\usepackage{mathrsfs}

\begin{document}

\title{Formation of Isotopically Heterogeneous Molecular Cloud Cores in Filamentary Molecular Clouds}

\author[0000-0002-4093-6925]{Yoshiaki Misugi}
\affiliation{Faculty of Science and Engineering, Kyushu Sangyo University, 2-3-1 Matsukadai, Fukuoka 813-8503, Japan}
\affiliation{Division of Science, National Astronomical Observatory of Japan, Osawa, Mitaka, Tokyo 181-8588, Japan}

\author[0000-0003-4366-6518]{Shu-ichiro Inutsuka}
\affiliation{Department of Physics, Graduate School of Science, Nagoya University, Nagoya, Aichi 464-8692, Japan}

\author[0000-0003-3924-6174]{Taishi Nakamoto}
\affiliation{Department of Earth and Planetary Sciences, Institute of Science Tokyo, Ookayama, Meguro, Tokyo 152-8551, Japan}


\author[0000-0003-2247-7232]{Tetsuya Yokoyama}
\affiliation{Department of Earth and Planetary Sciences, Institute of Science Tokyo, Ookayama, Meguro, Tokyo 152-8551, Japan}

\begin{abstract}
Meteorite analysis shows that the older solids of the solar system, such as the calcium–aluminum-rich inclusions (CAIs), have isotopic inhomogeneity. This indicates that the isotopic inhomogeneity could originate from parental molecular clouds.   We investigate the evolution of the isotopically heterogeneous molecular cloud cores formed from filament fragmentation using the smoothed particle hydrodynamics method. We show that the effect of the variation of isotopic ratio along the minor axes of the filament is smaller than that along the longitudinal axis of the filament due to the filament geometry. Our results also suggest that isotopic inhomogeneities remain in the resulting cores, although the amounts of initial inhomogeneities are reduced by a factor of $\sim 100$ from those over the initial filament length of 1 pc. This fraction corresponds to 1-10\% of the maximum isotopic ratio that the core can acquire from the filament in each model. The origin of the isotopic inhomogeneity of the shells could be attributed to the initial difference in the center-of-mass of shells caused by the turbulent velocity field. Our model indicates that the isotopic inhomogeneity could survive even in the circumstellar disk.
\end{abstract}

\keywords{Interstellar filaments(842) --- Meteorites(1038) ---  Protoplanetary disks(1300) --- Star formation(1569) }

\section{Introduction} \label{sec:intro}

Young stellar objects are supposed to form in the dense molecular cloud cores whose typical size is on the order of 0.1 pc \citep[e.g., ][PPVII]{Pineda2023}.  
Recent observations have established the scenario that dense molecular cores are created by the self-gravitational fragmentation of dense filamentary molecular clouds whose mass per unit length is greater than the critical line-mass ($ 2 c_{\rm s}^{2} / G $, where $c_{\rm s}$ is the sound speed and $G$ is the gravitational constant, \cite{Andre2014}). 
This scenario is observationally established, at least, in the Solar neighborhood. 
The formation of dense filamentary clouds has been extensively studied, and \cite{Abe2021} categorized the formation mechanisms into six different processes. Among them, the mechanism of the oblique magneto-hydrodynamical shock wave seems to explain most of the observational findings. 
Further investigation in this line is now providing the reason why massive filaments tend to keep the width on the order of 0.1 pc for a long time, which was identified by the observations \citep{Arzoumanian_2011}, but has been a puzzling question for theorists. 
The magneto-hydrodynamical shock waves should be the main compression mechanism of the magnetized interstellar medium, which can create HI clouds and molecular clouds from a warm neutral medium. 
This process has also been extensively investigated, and the difficulty in creating a magnetized molecular cloud in a single compressional event has been identified by many authors \citep{vanLoo2007,Inoue2008,Inoue2009,Inoue2012,Heitsch2009,Kortgen2015,Valdivia2016, Iwasaki2019}. 
The current conclusion is that the formation of magnetized molecular clouds requires multiple episodes of compressional events, mostly due to old expanding supernova remnants whose shock velocity is on the order of 20 km/s \citep{Inutsuka2015}.

In general, expanding supernova remnants spread and mix heavy elements created inside massive stars towards the surrounding interstellar medium. 
This process leads to the injection of various isotopes of heavy elements into the molecular clouds. The injection of elements from supernova remnants is thought to have been recorded in the dust particles that were present in the early Solar System. The vast majority of meteorites found and collected on Earth originate from the objects within the asteroid belt, located between the orbits of Mars and Jupiter. These asteroids are believed to be remnants of planetesimals and protoplanets that accreted from dust particles in the protoplanetary disk. Some of these bodies remained nearly primitive, undergoing no significant changes such as melting (e.g., carbonaceous chondrites). A comprehensive investigation into the isotopic compositions of these primitive meteorites can provide insights into the origin of materials present in the early Solar System. 
In this process, various isotopes of heavy elements can be injected into the molecular clouds. 
For instance, the amount of a short-lived iron isotope $^{60}$Fe ($T_{1/2} = 2.62 \times 10^6$ yr) into the parental molecular cloud was theoretically studied in relation to the initial $^{60}$Fe$/^{56}$Fe ratio of the Solar System, as constrained by the measurement of meteorites \citep{Dauphas2008, Gounelle2009}.

The composition of stable isotopes in meteorites has been extensively studied, which has revealed variations in the isotopic ratios of certain elements among different meteorites \citep{Dauphas2016}.
For instance, the isotopic ratios of $^{54}$Cr$/^{52}$Cr and $^{50}$Ti$/^{47}$Ti exhibit relative differences on the order of 1 to 10 parts in 10,000 across meteorites \citep{Trinquier2009}.
The origin of this isotopic heterogeneity has been a subject of extensive discussion. 
One proposed explanation is the thermal processing model, which suggests that dust particles in the protoplanetary disk underwent thermal processing, leading to partial vaporization and chemical differentiation \citep{Trinquier2009, Ek2020}.
Nuclides produced in nucleosynthetic sites are incorporated into dust grains, each with differing resistance to thermal processing.
Since the temperature in the protoplanetary disk varies with both time and location, dust grains that experienced partial vaporization would retain different amounts of nuclides depending on their thermal resilience, while the model remains poorly tested \citep{Dauphas2016}.
An alternative explanation is the inhomogeneous parent molecular cloud core model \citep{Dauphas2002}, which posits that isotopic heterogeneity was inherited from the molecular cloud core that formed the protoplanetary disk.
In support of this scenario, \cite{Jacquet2019} investigated the distribution of isotopes in a disk formed by the gravitational collapse of a molecular cloud core and found that the degree of isotopic heterogeneity in the resulting disk was reduced by a factor of three relative to that in the initial molecular cloud core.
This finding suggests that, under certain conditions, the observed isotopic heterogeneity in Solar System materials may be naturally explained by the presence of an appropriate level of inhomogeneity in the parental molecular cloud core.\par
In the inhomogeneous parent molecular cloud core model, the subsequent key inquiry pertains to the extent of isotopic heterogeneity present in the filamentary molecular clouds that give rise to molecular cloud cores.
This inhomogeneity may be shaped by the history of cloud formation, the spatial and temporal proximity to nucleosynthetic sources, and other environmental factors.
Examining the isotopic characteristics of the filamentary molecular cloud that formed the Solar System can thus provide valuable insights into the astrophysical birth environment.
To this end, we examine the gravitational fragmentation of a filamentary molecular cloud that initially contained isotopic inhomogeneity and follow the evolution of this inhomogeneity during core formation.
This approach enables estimation of the efficiency with which the initial inhomogeneity is preserved or smoothed out through the collapse process.
By integrating this efficiency with the reduction factor reported by \cite{Jacquet2019} and the isotopic heterogeneity observed in Solar System materials, we infer the degree of inhomogeneity in the filamentary molecular cloud that gave rise to the Solar System. \par

This paper is organized as follows. Section \ref{sec:setup} details the numerical method and the setup. In Section \ref{sec:res}, we describe the isotopic inhomogeneity in the cores measured in our simulations. The isotopic inhomogeneity in the circumstellar disk is also discussed in Section \ref{sec:dis}. Finally, we summarize our results in Section \ref{sec:sum}.

\section{Numerical Setup} \label{sec:setup}
To investigate the evolution of isotopically heterogeneous molecular cloud cores, we analyze the results of simulations performed in \cite{Misugi2024}. Since the detail of the setup of our simulation is described in \cite{Misugi2024}, here, we briefly explain the setup of the simulation. We solve the ideal magnetohydrodynamic equations using a Godunov smoothed particle magnetohydrodynamic (GSPM) method \citep[][K. Iwasaki et al. 2026, in preparation]{Iwasaki2011,Iwasaki2013}. We adopt the periodic boundary condition along the $z$-direction, which corresponds to the direction of the filament's longitudinal axis. The boundaries of the $x$ and $y$ directions are far enough that the Alfv\'en wave does not reach the boundary in computational time. We parallelized our code with the Framework for Developing Particle Simulator \cite[FDPS, ][]{Iwasawa2016}. The mass of an SPH particle is $2.5 \times 10^{-5} \ M_{\odot}$ in our simulations. The initial filament consists of about $2\times 10^6$ SPH particles.\par
The initial density profile of the filament follows the hydrostatic equilibrium profile:
\begin{eqnarray}
\label{eq:inidenpro}
\rho(\xi)=\rho_{\mathrm{c} 0}\left[1+\left(\frac{\xi}{H_{0}}\right)^{2}\right]^{-2},
\end{eqnarray}
where $\rho_{\rm c0}$ is the initial density at the ridge of the filament, $\xi$ denotes the radius in the cylindrical coordinate system, and $H_{0}$ represents the scale height of the filament:
\begin{eqnarray}
H_{0} \equiv \sqrt{\frac{2 c_{\rm s}^{2}}{\pi {G} \rho_{\rm c0}}}.
\label{eq:scaleheight}
\end{eqnarray}
Here, $c_{\rm s}=0.2\  {\rm km \ s^{-1}}$ denotes the sound speed for $T=10 \ {\rm K}$. We choose $H_0=0.05\ {\rm pc}$ since the observations show that the filament width is universal and the distribution of the width has a sharp peak around 0.1 {\rm pc} \citep{Arzoumanian_2011,Arzoumanian2019}. The filament of the simulation has the critical line mass, $ {\rm M_{\rm line,crit}}=18 \ {\rm M_{\odot} pc^{-1}}$ for $T=10$ K. We adopt the Kolmogorov turbulence as the initial turbulent velocity field \citep{Misugi2019}, and the initial velocity dispersion is $\sigma=2c_s$. We perform the simulations with the strength of the initial magnetic field of $10 {\rm \mu G}$. The direction of the initial magnetic field is perpendicular to the filament axis and along the $x$-axis. In addition, we also perform the hydrodynamics simulation to study the effect of the magnetic field. We run 40 sets of simulations for the hydrodynamical case and 40 sets of simulations for the case with $B_0=10 {\rm \mu G}$ using different turbulent seeds in each parameter.\par
We construct six models to set up the initial profile of isotopic inhomogeneity. The details of the models are summarized in Table \ref{tab:model}. For example, in the model M\_z, the isotopic ratio variation is only along the $z$-axis (the filament axis). On the other hand, the model M\_xyz has the isotopic ratio variation along all directions: the $x$, $y$, and $z$-axes. As in \cite{Jacquet2019}, the isotopic ratio is normalized in the range from 0 to 1 in all models. Because the equation for the evolution of isotopic ratio is linear, the normalized evolution does not change by rescaling the initial isotopic ratio. Therefore, the degree of normalized isotopic heterogeneity obtained in our simulations is independent of the initial value of isotopic ratio. We label the SPH particles using the function shown in Table \ref{tab:model} at the initial state of our simulations. Then, we investigate the evolution of the isotopic ratio in the cores by tracing the SPH particles. \par
In the present model, isotopic species are treated as passive tracers in the simulations. Consequently, the isotopic ratio assigned to each SPH particle remains unchanged with time. This treatment neglects chemical processes that modify isotopic ratios, such as differential photodissociation and chemical fractionation. These isotope-selective processes are generally important for volatile molecular species containing elements such as oxygen and sulfur. In contrast, chromium is a refractory element and is expected to be incorporated into solid dust grains in molecular clouds. Therefore, isotope-selective chemical effects are expected to be negligible in the context considered here.

\begin{deluxetable*}{cc}
\tablenum{1}
\tablecaption{List of the models. \label{tab:model}}
\tablewidth{0pt}
\tablehead{
\colhead{Name of model} & \colhead{Function form of the initial isotopic ratio}
}
\startdata
M\_z & $C_0(z) = z/1.6\ {\rm pc}$\\
M\_x & $C_0(x) = (x+0.8\ {\rm pc})/1.6\ {\rm pc}$\\
M\_y & $C_0(y) = (y+0.8\ {\rm pc})/1.6\ {\rm pc}$\\
M\_xz & $C_0(x,z) = [z/1.6\ {\rm pc} +  (x+0.8\ {\rm pc})/1.6 \ {\rm pc}]/2$\\
M\_yz & $C_0(y,z) = [z/1.6\ {\rm pc} +  (y+0.8\ {\rm pc})/1.6 \ {\rm pc}]/2$\\
M\_xyz & $C_0(x,y,z) = [(x+0.8\ {\rm pc})/1.6 \ {\rm pc} + (y+0.8\ {\rm pc})/1.6 \ {\rm pc} + z/1.6\ {\rm pc}]/3$\\
\enddata
\tablecomments{The isotopic ratio is defined in the range of $x$=[$-0.8$ pc, 0.8 pc], $y$=[$-0.8$ pc, 0.8 pc], and $z$=[0 pc, 1.6 pc]. The isotopic ratio is normalized in [0, 1] in the domain.}
\end{deluxetable*}

\section{Results} \label{sec:res}

\subsection{Overview\label{subsec:obv}}

\begin{figure*}
\gridline{\fig{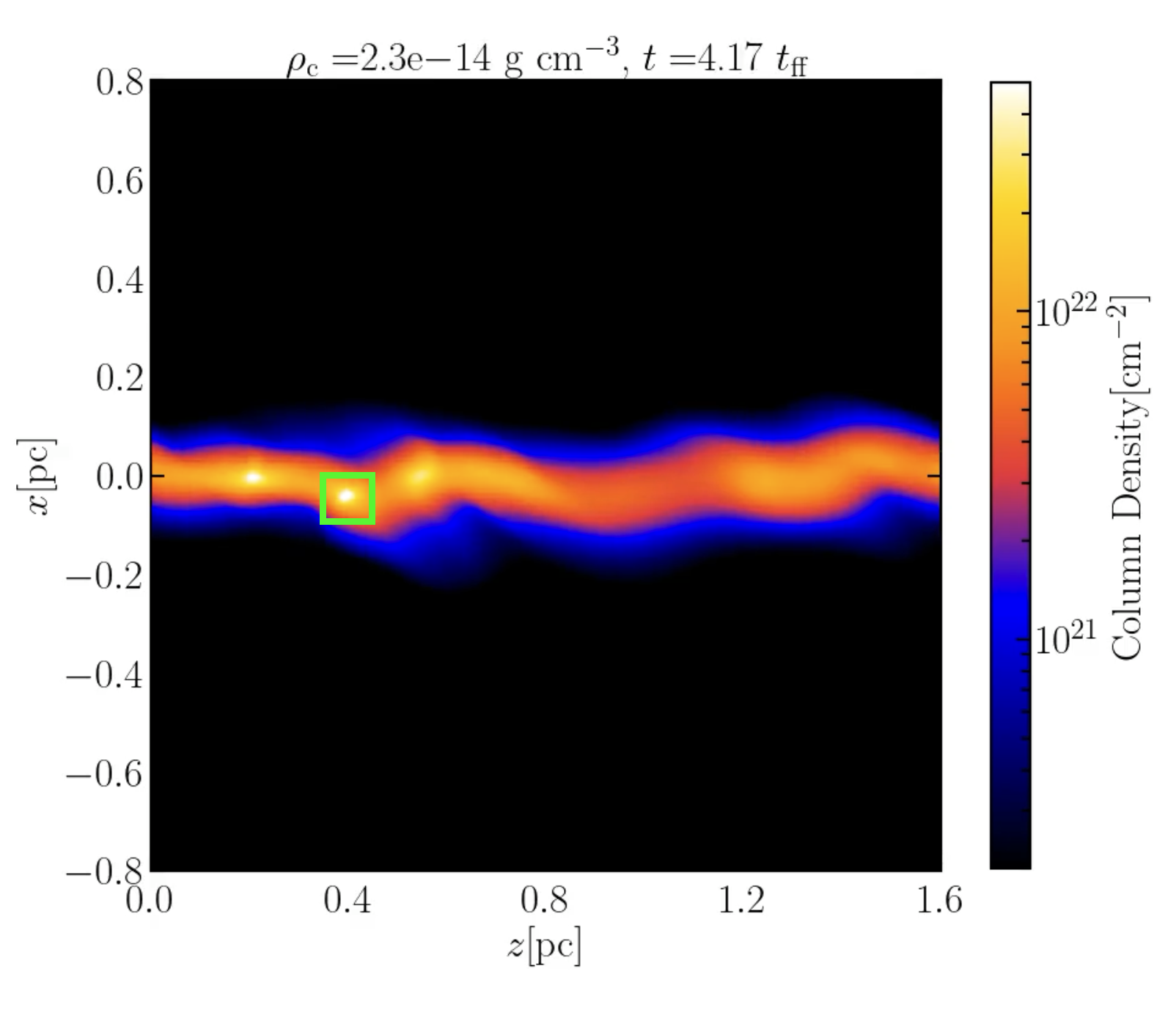}{0.4\textwidth}{(a)}}
\gridline{\fig{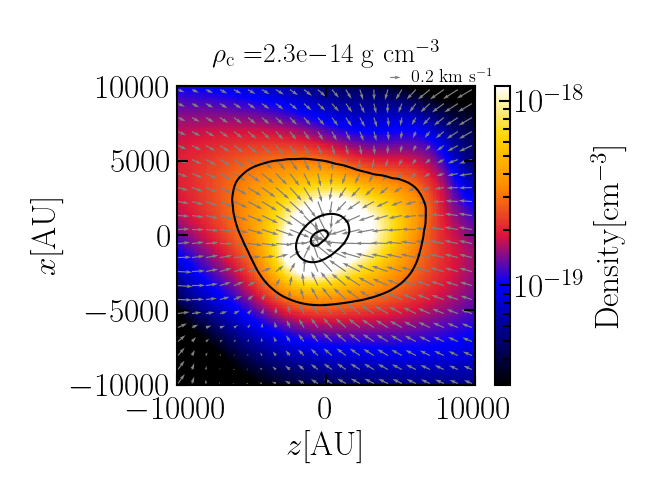}{0.45\textwidth}{(b)}
          \fig{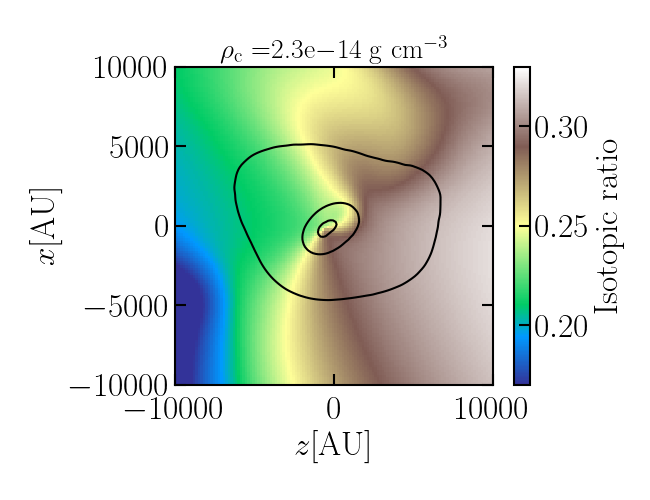}{0.45\textwidth}{(c)}}
\gridline{\fig{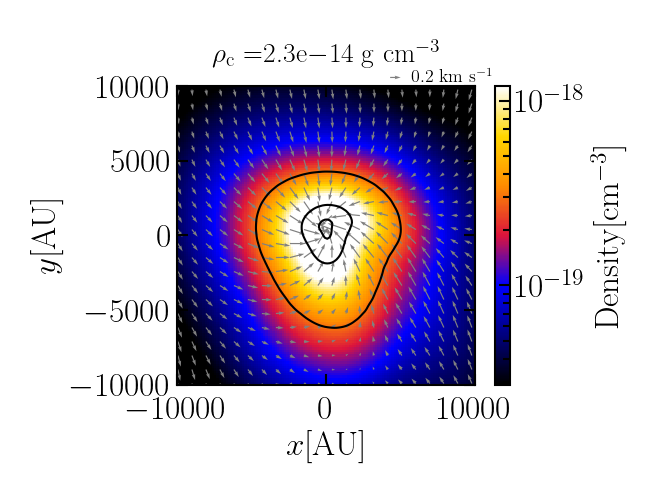}{0.45\textwidth}{(d)}
          \fig{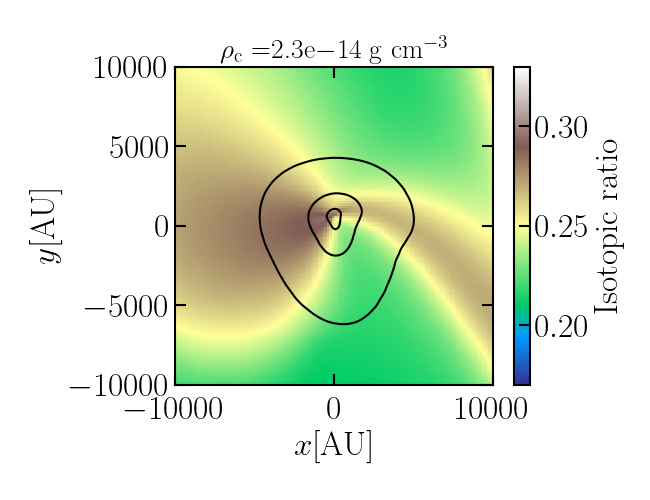}{0.45\textwidth}{(e)}}

\caption{Distribution of the density (Figure \ref{fig:overview} (b) and (d)) and isotopic ratio (Figure \ref{fig:overview} (c) and (e)) around the core enclosed by the green square in Figure \ref{fig:overview} (a). Figure \ref{fig:overview} (a) shows the column density map of the filament at the final state of our simulation. Figure \ref{fig:overview} (b) and (c) are the slice maps on the $z$-$x$ plane. Figure \ref{fig:overview} (d) and (e) are the slice maps on the $x$-$y$ plane. The contour levels in Figure \ref{fig:overview} (b) - (e) represent $\rho = 3 \times 10^{-19} \ {\rm g \ cm^{-3}}$, $\rho = 3 \times 10^{-18} \ {\rm g \ cm^{-3}}$, and $\rho = 3 \times 10^{-17} \ {\rm g \ cm^{-3}}$.
 The gray arrows represent the velocity vectors. In Figure \ref{fig:overview} (c) and (e), we adopt the model M\_z (Table \ref{tab:model}). \label{fig:overview}}
\end{figure*}

Figure \ref{fig:overview} (a) displays the column density map of the filament in the hydrodynamic case at the final state of the simulation. The filament fragments into the cores because of the growth of the fluctuations of the initial turbulence. In Figure \ref{fig:overview} (c) and (e), we show the resultant isotopic distribution around the core in the case of the model M\_z. Since we adopt the model M\_z, the initial isotopic ratio variation is along the $z$-axis, and there is no isotopic ratio variation on the $x$-$y$ plane at the initial state. However, because of the initial turbulent velocity field, the non-axisymmetric distributions of the isotopic ratio appear in Figure \ref{fig:overview} (c) and (e). As can be seen from the velocity field in Figure \ref{fig:overview} (b) and (d), the convergent accretion flow is related to the non-axisymmetric distributions of the isotopic ratio. For example, around $(z,x) \sim (5000\ {\rm AU},\ 5000\ {\rm AU} )$ in Figure \ref{fig:overview} (b), the flow that comes from the left meets the flow that comes from the opposite direction. These flows bring gas fluids that have different isotopic ratios, and mixing could occur at the intersection of the flows. This flow pattern would create the inhomogeneity of the isotopic ratio, as can be seen around $(z,x) \sim (5000\ {\rm AU},\ 5000\ {\rm AU} )$ in Figure \ref{fig:overview} (c). We investigate the structure of the isotopic ratio in the cores in more detail in the following section.

\subsection{Internal Profile of the Isotopic Ratio\label{subsec:modelz}}

\begin{figure}[t]
\plotone{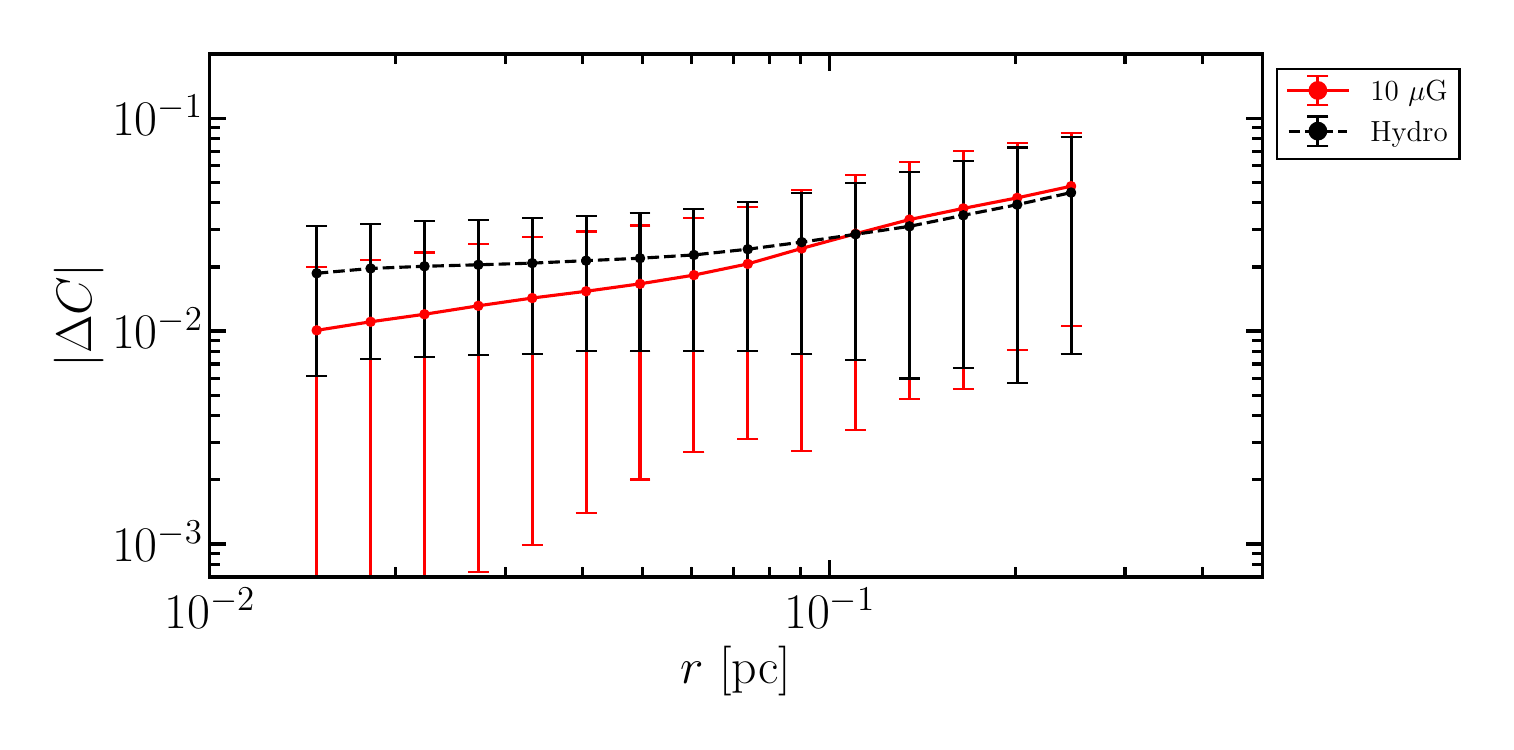}
\caption{Internal profile of the isotopic ratio in the case of the model M\_z. The vertical axis is the difference between the average isotopic ratio of the shell and that of the core center (see Equation \ref{eq:delc_def}). The horizontal axis is the distance from the density peak. The red and black solid lines are the profile in the case of the hydrodynamics and $B=10 \ {\rm \mu G }$, respectively. The profiles are averaged over 40 cores identified in our simulations.\label{fig:Cprofile}}
\end{figure}

In this subsection, first, we focus on the model M\_z. To investigate the internal profile of the isotopic ratio, we define the spherical shells around the density peak of the core and calculate the isotopic ratio of the shells as follows:
\begin{eqnarray}
\bar{C}(r)=\frac{\int\int C(r, \theta, \varphi) \rho(r, \theta, \varphi) \sin \theta d \theta d \varphi}{\int\int \rho(r, \theta, \varphi) \sin \theta d \theta d \varphi}.
\label{eq:shell}
\end{eqnarray}
Here $\rho(r, \theta, \varphi)$ is the density profile, and the origin of the coordinate is the density peak of the core. The inhomogeneity along the $\theta$ and $\phi$ directions is expected to be smeared out due to the shear motion in protoplanetary disks once the disk forms around the protostar. In that sense, we take the average over the $\theta$ and $\phi$ directions in Equation \ref{eq:shell}. For a more realistic case, in Section \ref{subsec:disk}, we construct a model taking into account the distribution of the angular momentum in the core to estimate the isotopic inhomogeneity in the circumstellar disks.\par
Figure \ref{fig:Cprofile} shows the internal profile of the isotopic ratio measured in the final state of the simulations in the case of the model M\_z. $|\Delta C(r)|$ is defined as;
\begin{eqnarray}
|\Delta C(r)| = |\bar{C}(r) - \bar{C}(r_{\rm fin, ref})|, 
\label{eq:delc_def}
\end{eqnarray}
where $r_{\rm fin, ref}=10^{-4} \ {\rm pc}$. We confirmed that $|\Delta C(r)|$ does not change even if we choose a smaller $r_{\rm fin, ref}$. The profile is averaged over all 40 cores. Figure \ref{fig:Cprofile} shows that the resultant profile of the isotopic ratio does not strongly depend on whether the initial filament is magnetized or not, since the resultant profiles do not have a large difference, taking into account the error bars. Figure \ref{fig:Cprofile} also indicates that the shell with a typical core radius of 0.05 pc has an isotopic ratio $|\Delta C| \sim 0.01$. This means that the isotopic ratio of the cores with a radius of 0.05 pc is $1\%$ of the difference of the isotopic ratio of the entire filament since the initial distribution of the isotopic ratio is normalized in [0, 1]. Since the mean value of the isotopic ratio in the computational domain is ⟨C⟩ = 0.5 among all models, the relative variation is simply given by $\Delta C/\left<C \right> = 2\Delta C$.

\subsection{Effect of the Isotopic Ratio Variation along the Minor Axis of the Filament\label{subsec:modelxy}}

\begin{figure}[t]
\plotone{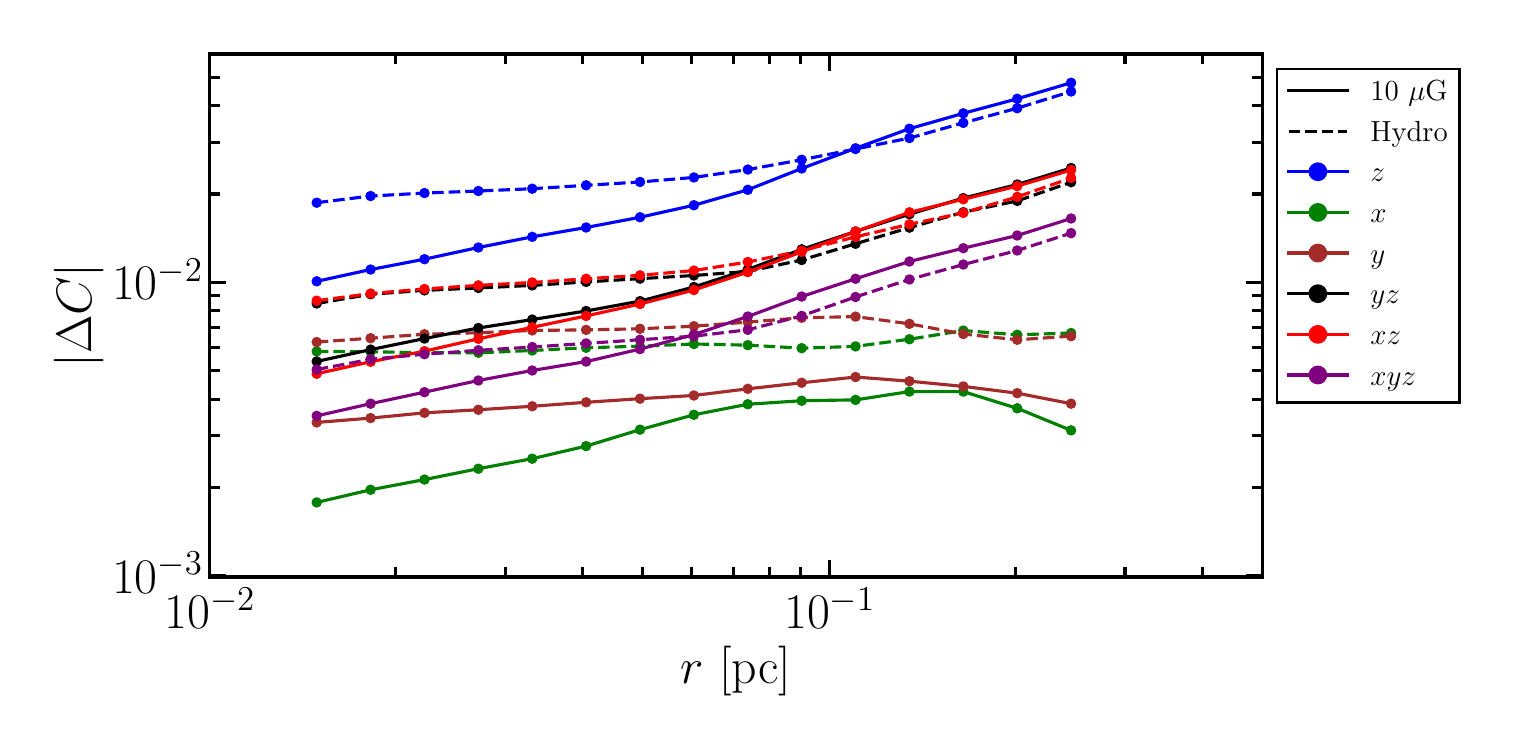}
\caption{Comparison of the internal profile of the isotopic ratio. The blue, green, and brown lines are the results of the models of M\_z,  M\_x, and M\_y, respectively (see Table \ref{tab:model}). The black, red, and magenta lines are the results of the models of M\_yz,  M\_xz, and M\_xyz, respectively. The solid and dashed lines represent the $B_0=10{\rm \mu G}$ and hydrodynamic cases, respectively.  The profiles are averaged over 40 cores found in our simulations.\label{fig:Cprofile_model}}
\end{figure}

\begin{figure*}
\gridline{\fig{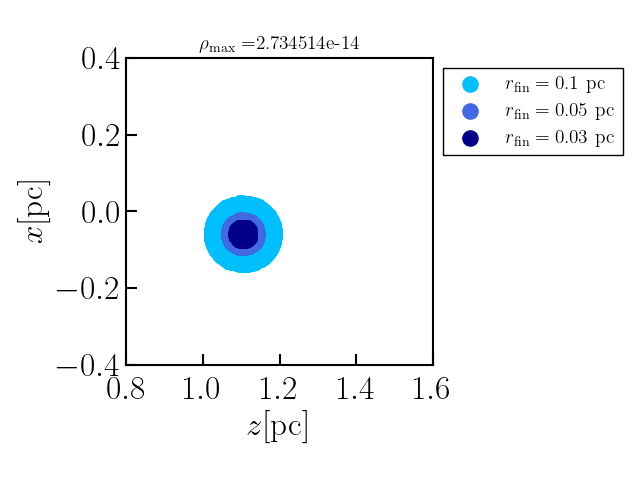}{0.48\textwidth}{(a)}
          \fig{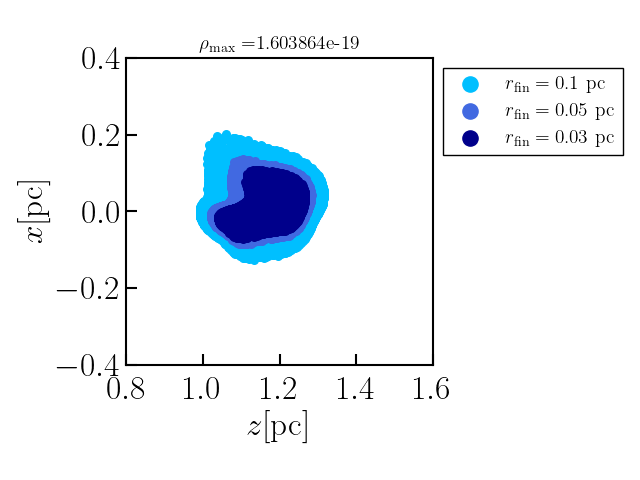}{0.48\textwidth}{(b)}}
\caption{SPH particles of each shell at the final state (a) and at the initial state (b). The sizes of the shells are 0.1, 0.05, and 0.03 pc from the outer to the inner regions. \label{fig:lag}}
\end{figure*}

\begin{figure}[t]
\plotone{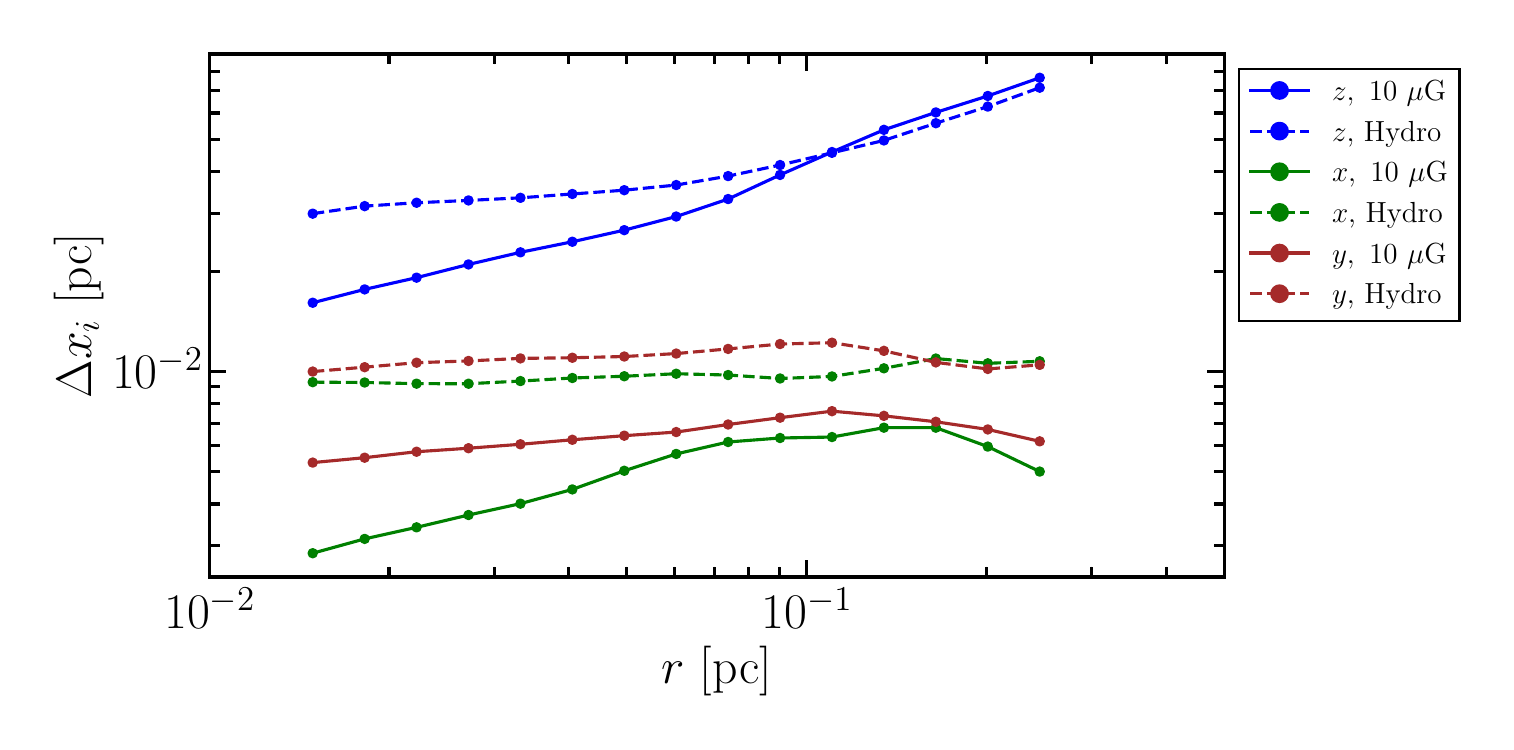}
\caption{Difference of the center-of-mass of each shell at the initial state. The blue, green, and brown lines represent the difference of the center-of-mass along the $z$, $x$, and $y$-axes, respectively. The solid and dashed lines are the results of the $B_0=10{\rm \mu G}$ and hydrodynamic cases, respectively.\label{fig:center}}
\end{figure}

\begin{figure*}
\gridline{\fig{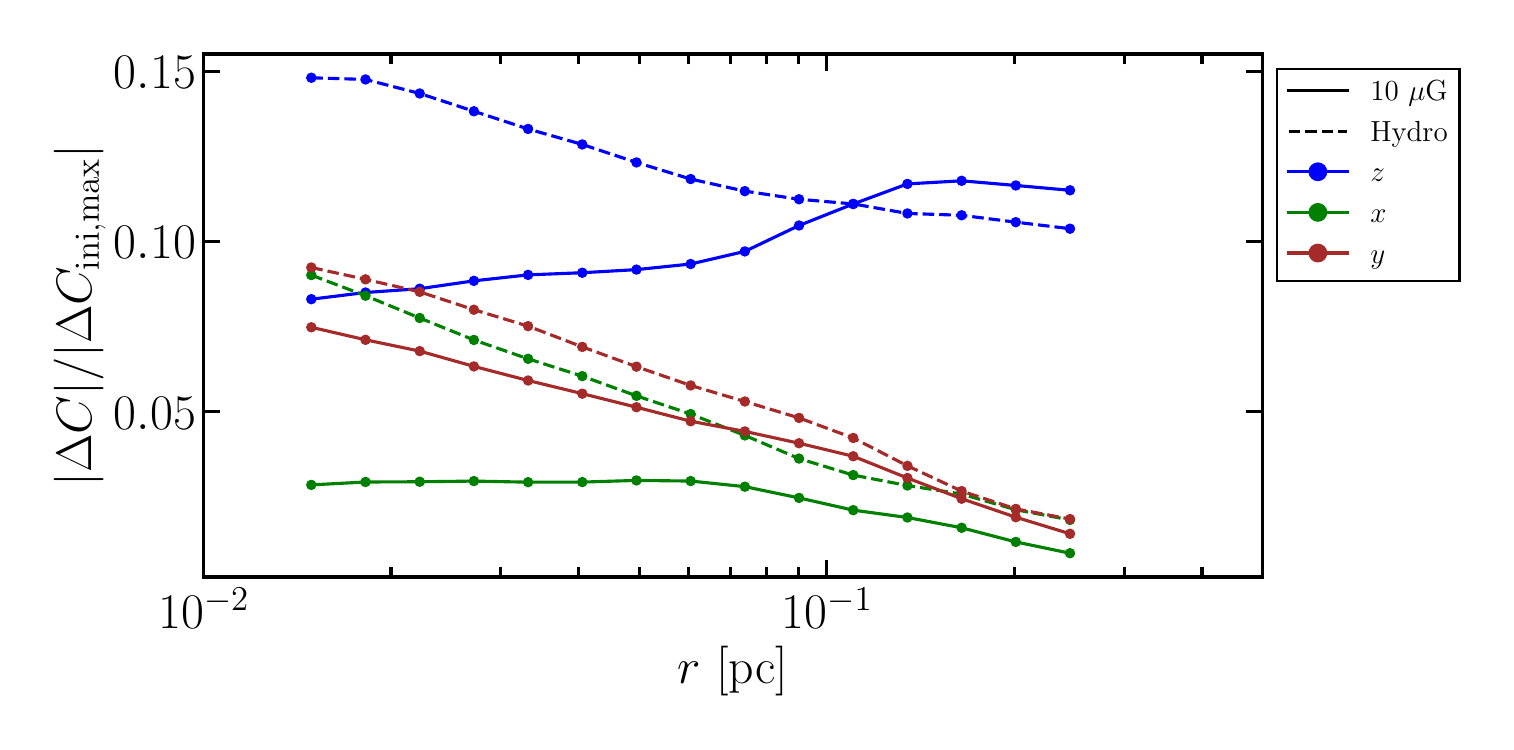}{0.8\textwidth}{(a)}}
\gridline{\fig{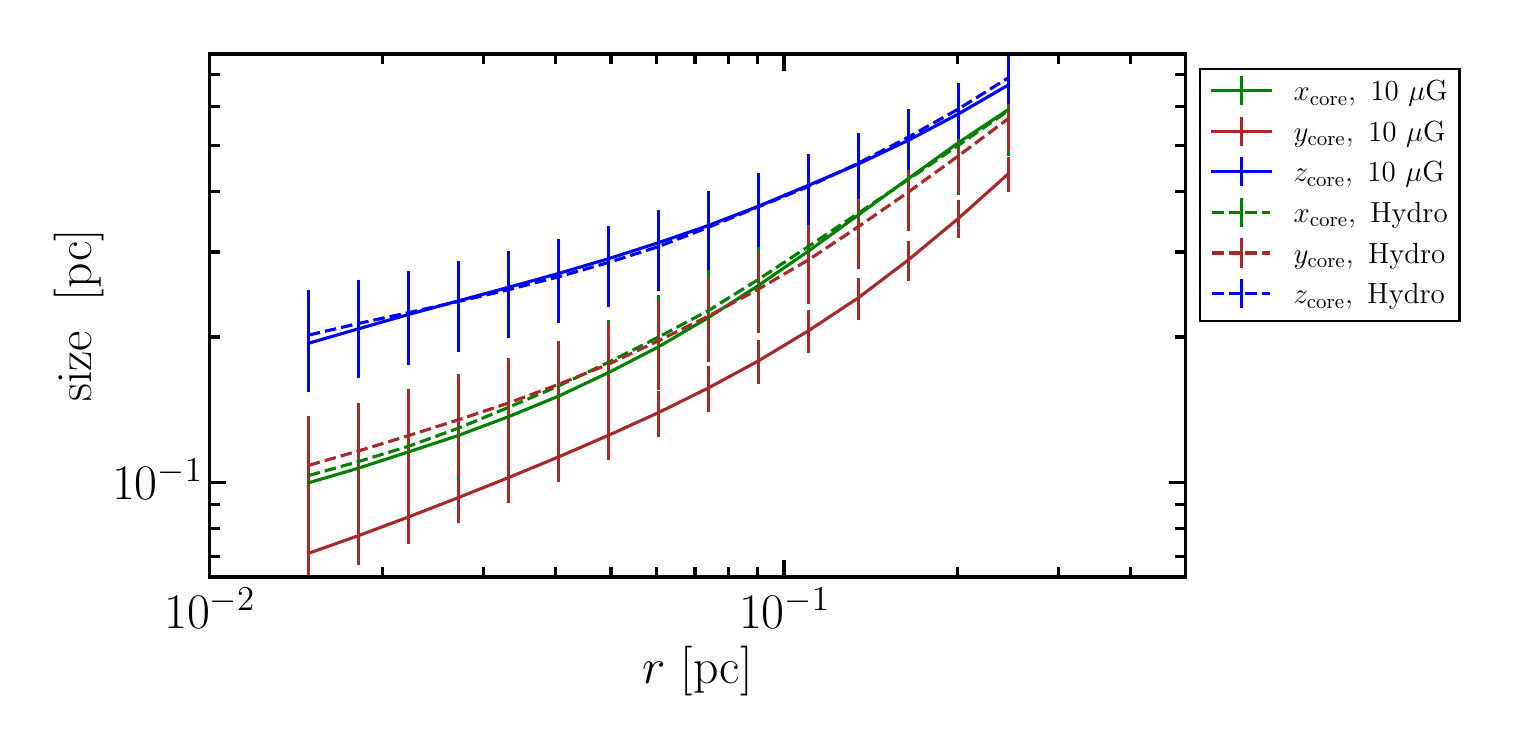}{0.8\textwidth}{(b)}}
\caption{(a) Isotopic ratio at the final state normalized by the maximum isotopic ratio that the shell can acquire at the initial state. The blue, green, and brown lines are the results of the models M\_z, M\_x, and M\_y, respectively. $|\Delta C_{\rm ini, max}|$ is estimated using the shell size shown in panel (b). The blue, green, and brown lines in panel (b) represent the shell size along the $z$-, $x$-, and $y$-axes, respectively. For example, the shell size along the $z$-axis is defined as $z_{\rm max}-z_{\rm min}$. Here, $z_{\rm max}$ and $z_{\rm min}$ are the maximum and minimum values of $z$-coordinate of the SPH particles contained in the shell. \label{fig:rate}}
\end{figure*}

In Section \ref{subsec:modelz}, we focused on the results of the model M\_z. In this subsection, we discuss the effect of the isotopic ratio variation along the minor axes of the filament. Figure \ref{fig:Cprofile_model} shows the dependence of the radial profile of the isotopic ratio on the model. First, we compare the profiles of model M\_z (the blue lines) to those of the models M\_x (the green lines) and M\_y (the brown lines). In Figure \ref{fig:Cprofile_model}, the isotopic ratio of the model M\_x and M\_y is smaller than that of the model M\_z. To explain this behavior, we need to understand the origin of the difference in the isotopic ratio of the shells. In our models, since we assume that the initial profile of the isotopic ratio has a linear gradient, the initial positions of the center-of-mass of the shells defined at the final state are closely related to the difference in the isotopic ratio between shells (see Equation \ref{eq:shell}). An example of the shells is illustrated in Figure \ref{fig:lag}. The left panel of Figure \ref{fig:lag} displays the shells defined at the final state of the simulation. Then, we trace the trajectories of SPH particles and plot each shell at the initial state in the right panel of Figure \ref{fig:lag}. Figure \ref{fig:lag} shows that the shape of the shells is different and complex, depending on each shell. Here, to quantify the effect of the difference in the initial position of fluid elements on the average isotopic ratio, we measure the position of the center-of-mass of each shell at the initial state of the simulations, and the results are shown in Figure \ref{fig:center}. Figure \ref{fig:center} shows that the difference of the center-of-mass of the $z$-direction is larger than those of the $x$-direction and the $y$-direction. In our calculations, since the filament has a width of 0.1 pc, the filament geometry does not allow a large displacement along the minor axis directions (the $x$- and $y$-directions). For the model M\_z, the difference of the center-of-mass along the $z$-axis (the blue lines in Figure \ref{fig:center}) is essential to generate the isotopic inhomogeneity between shells. On the other hand, for the model M\_x, the isotopic inhomogeneity of the core is determined by the initial displacement along the $x$-direction (the green lines in Figure \ref{fig:center}). Therefore, the dependence of the isotopic inhomogeneity on the model is caused by the filament geometry. Note that, in the models M\_x and M\_y, the isotopic ratio becomes flat in the outer region due to the filament geometry. For the models M\_xz, M\_yz, and M\_xyz, the difference in the isotopic ratio is smaller than that of model M\_z since the isotopic ratio is normalized in the computational domain (Table \ref{tab:model}). The profiles of the models M\_xz, M\_yz, and M\_xyz do not become flat since the profile is mainly determined by the isotopic ratio variation along the $z$-direction. \par 
Figure \ref{fig:Cprofile} and \ref{fig:Cprofile_model} indicate that 0.1\% - 1\% of the initial isotopic inhomogeneity in the filament survives in the molecular cloud cores. However, it is meaningful to estimate what fraction of the initial isotopic inhomogeneity in the core survives. This is estimated by the ratio of the isotopic ratio measured at the final state, as shown in Figure \ref{fig:Cprofile} and \ref{fig:Cprofile_model}, to the isotopic ratio of the core defined at the initial state. The resultant ratio is shown in Figure \ref{fig:rate}(a). The vertical axis represents the difference of the isotopic ratio normalized by the maximum isotopic ratio which the core can acquire at the initial state, $|\Delta C_{\rm ini, max}|$. $|\Delta C_{\rm ini, max}|$ is estimated using the core size at the initial state shown in Figure \ref{fig:rate}(b). For example, if we adopt the model M\_z, we calculate $|\Delta C_{\rm ini, max}| = C_0(z)z_{\rm core}$, where $z_{\rm core}$ is a core size along the $z$-axis. On the other hand, in the case of the model M\_x, $|\Delta C_{\rm ini, max}|$ is estimated as $|\Delta C_{\rm ini, max}| = C_0(x)x_{\rm core}$. Figure \ref{fig:rate} indicates that 8\% - 15\% of the maximum inhomogeneity of the isotopic ratio at the initial state survives in the case model M\_z, and even in the case of the models M\_x and M\_y, 1\% - 10\% of the initial inhomogeneity of the core still exists. Note that the initial inhomogeneity estimated here is the maximum inhomogeneity because, to create the isotopic ratio variation $|\Delta C_{\rm ini, max}|$ in the core, the fluid element at the edge of the initial core needs to fall onto the center of the core and the fluid element at the other edge remain at the edge of the shell during the filament fragmentation phase. \\
\\

\section{Discussion} \label{sec:dis}

\subsection{Semi-Analytical Estimate of the Isotopic Inhomogeneity of Cores}\label{subsec:ana}

\begin{figure*}

\gridline{\fig{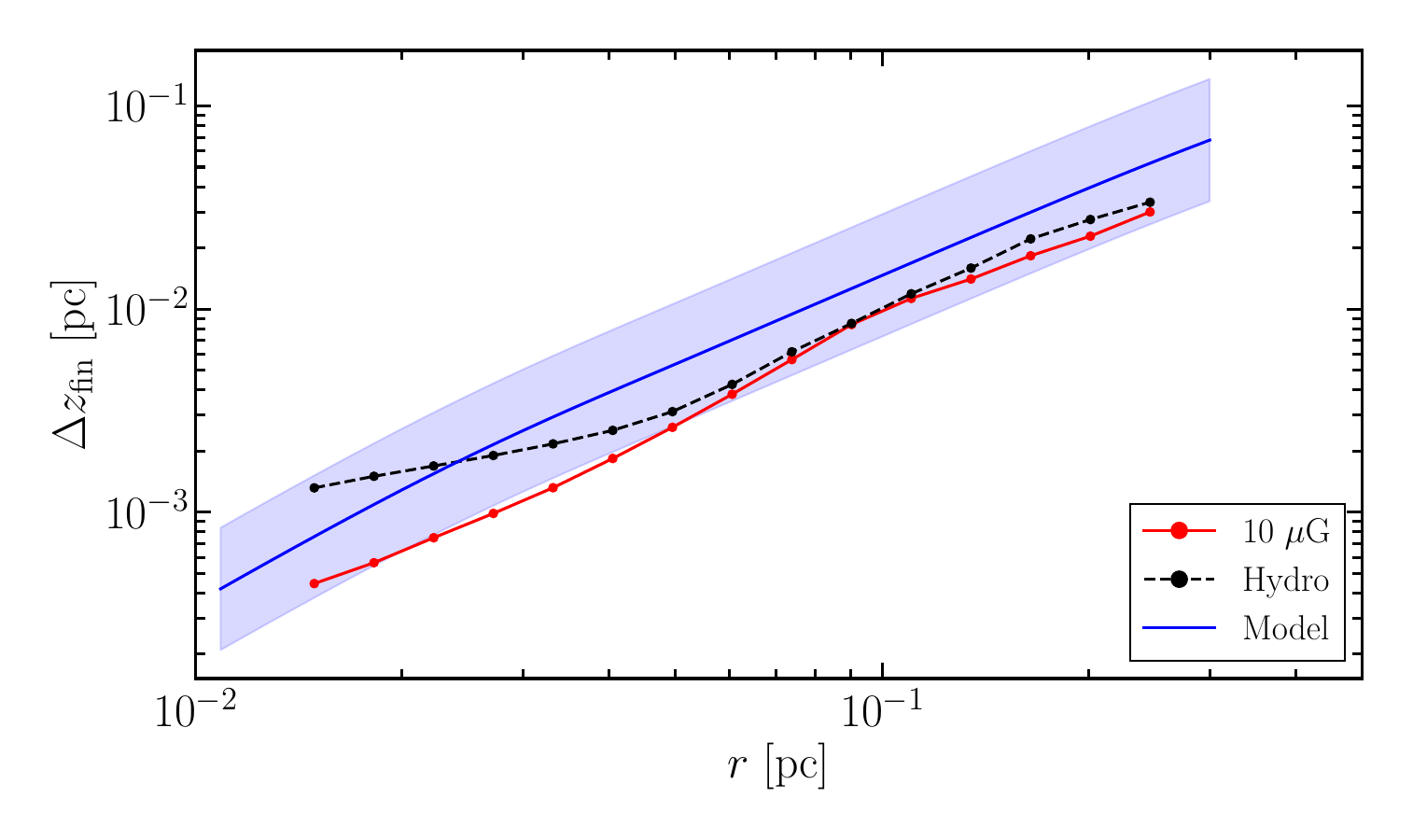}{0.48\textwidth}{(a)}
          \fig{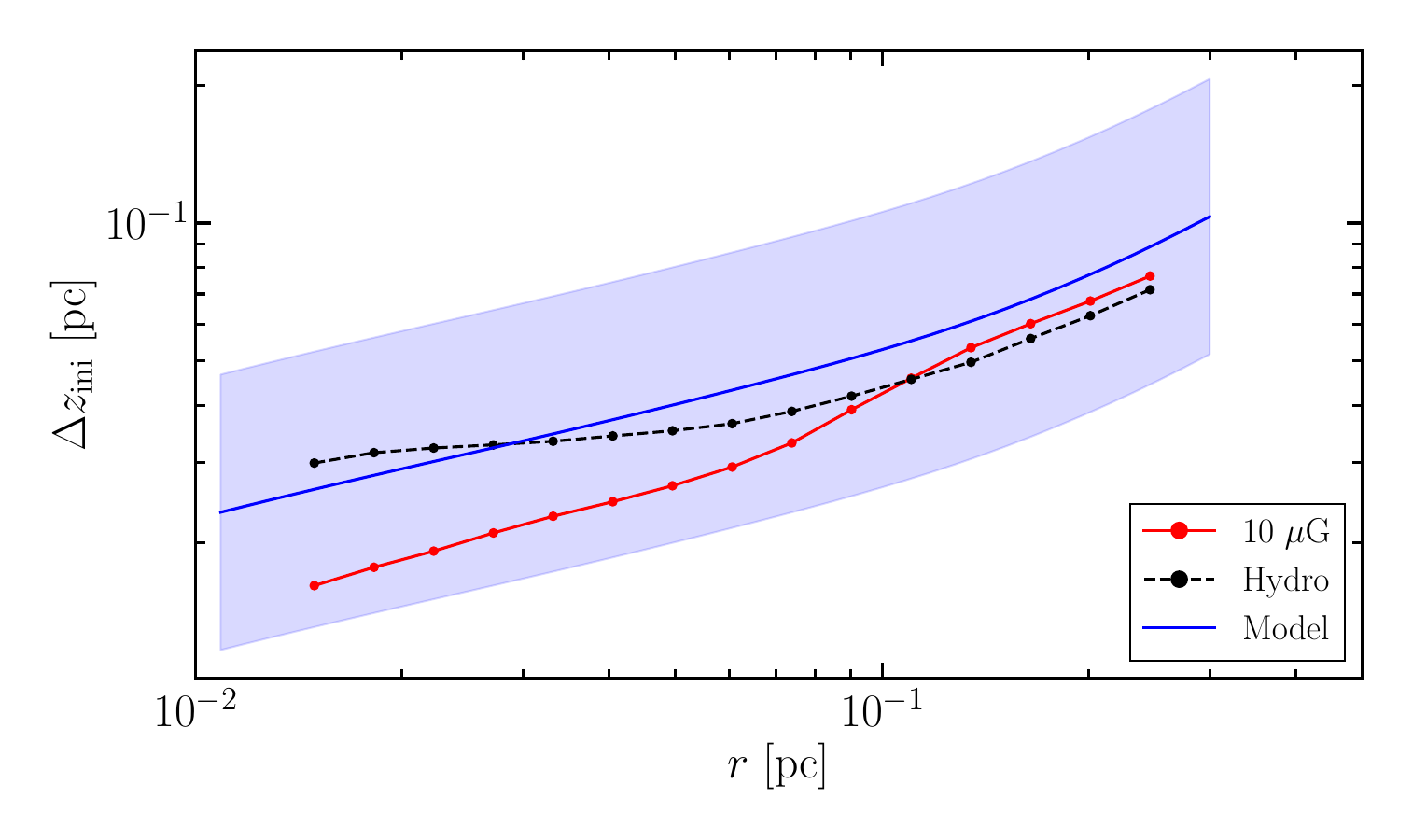}{0.48\textwidth}{(b)}}
\gridline{\fig{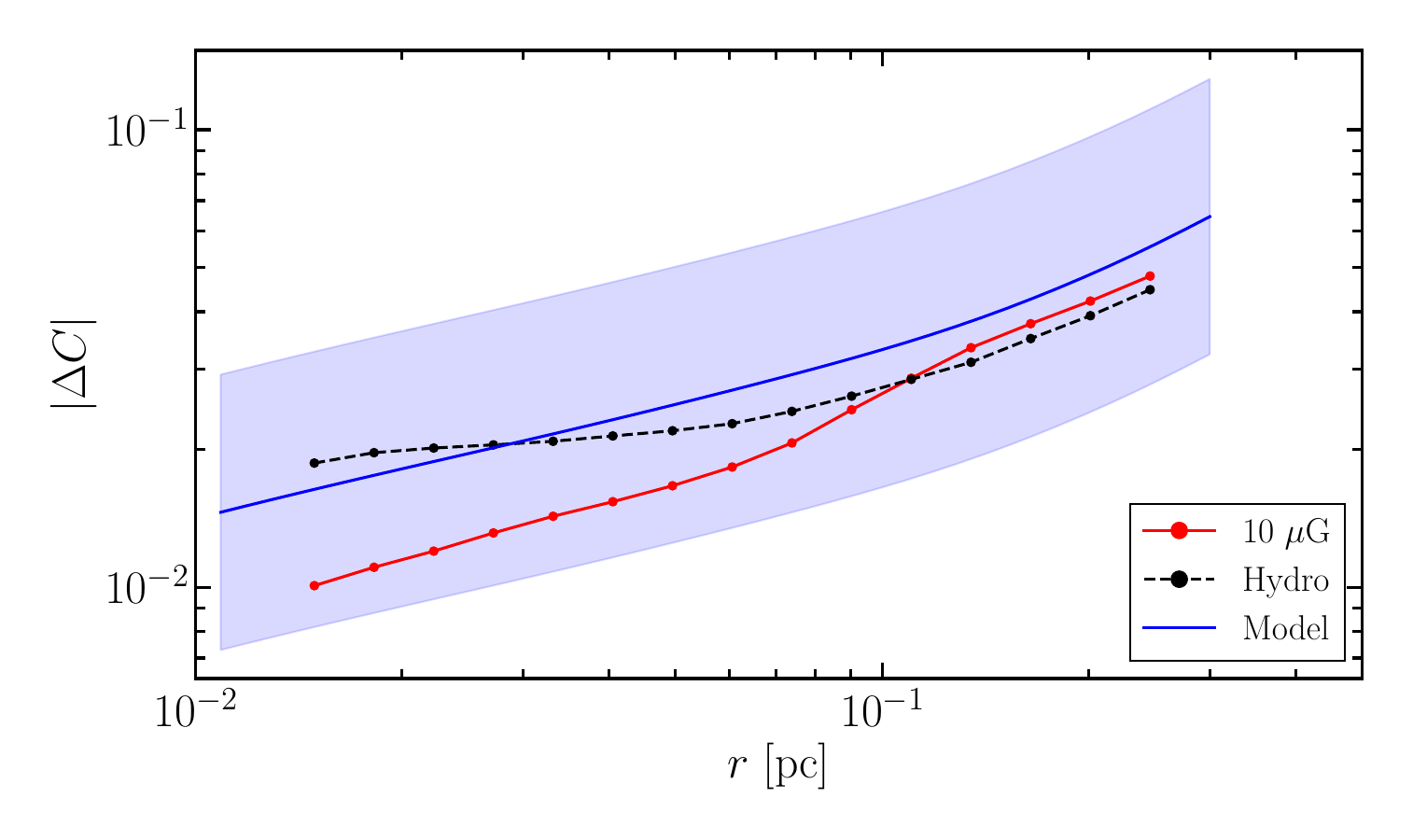}{0.48\textwidth}{(c)}}

\caption{Comparison of the semi-analytical estimate and simulation results. The blue solid line represents to $\Delta z_{\rm fin} \equiv \sqrt{\left< |\delta z_{\rm fin}|^2 \right>} $ (a), $ \Delta z_{\rm ini}  \equiv \sqrt{ \left< |\delta z_{\rm ini}|^2 \right>} $ (b), and $\Delta C \equiv \Delta z_{\rm ini}/1.6$ (c). The blue-shaded regions correspond to the regions within a factor of 2 of the semi-analytical model.
\label{fig:ana}}
\end{figure*}

In Section \ref{sec:res}, we discussed the isotopic ratio in the cores formed through filament fragmentation. We found that $\sim$ 1 \% of the initial inhomogeneity in the entire filament survives even just before the first core formation (Figure \ref{fig:Cprofile_model}). In this subsection, we estimate the isotopic inhomogeneity in the core quantitatively using a semi-analytical approach. As shown in Figure \ref{fig:Cprofile_model}, the isotopic ratio variation along the $z$-axis makes the largest impact on the resultant isotopic ratio. Therefore, in this subsection, we focus on the model M\_z. \par
First, the difference in the center-of-mass of shells at the initial state $\delta z_{\rm ini}$ can be written as follows:
\begin{eqnarray}
\delta z_{\rm ini} &\simeq \delta z_{\rm fin} + \delta v_{\rm fin} t \nonumber \\
&\simeq \delta z_{\rm fin} + \delta v_{\rm ini} t
\label{eq:delz}
\end{eqnarray}
$\delta z_{\rm fin}$ denotes the difference in the center-of-mass of shells at the final state. $\delta v_{\rm fin}$ is the difference in the velocity of the center-of-mass of shells at the final state. $\delta v_{\rm ini}$ and $t$ represent the difference in the velocity of the center-of-mass of shells at the initial state and the timescale of our simulations. For simplicity, we ignore the acceleration term. The ensemble average of Equation \ref{eq:delz} can be written as follows:
\begin{eqnarray}
\left< |\delta z_{\rm ini}|^2 \right> \simeq \left< |\delta z_{\rm fin}|^2 \right> + \left< |\delta v_{\rm ini}|^2 \right>t^2.
\label{eq:ens}
\end{eqnarray}
For simplicity, we ignore the cross term. To derive $\left< |\delta z_{\rm ini}|^2 \right>$, we need to estimate $\left< |\delta v_{\rm ini}|^2 \right>$ and $\left< |\delta z_{\rm fin}|^2 \right>$. 
To estimate $\left< |\delta z_{\rm fin}|^2 \right>$, we consider a spherical shell with a radius of $r_{\rm fin}$ as we did in the analysis of the simulations. If the core is spherical symmetry, $\left< |\delta z_{\rm fin}|^2 \right> = 0$. However, density fluctuations are generated by the turbulent velocity field in our simulations. These density fluctuations generate a finite shift of the center-of-mass, $\left< |\delta z_{\rm fin}|^2 \right> \neq 0$. $\left< |\delta z_{\rm fin}|^2 \right>$ can be estimated as follows:  

\begin{eqnarray}
\left< |\delta z_{\rm fin}|^2 \right> = \sum_{\bm k} P_{\rm den}(k) \left[ r_{\rm fin} j_{\rm B}^{(1)}(kr_{\rm fin}) - r_{\rm ref} j_{\rm B}^{(1)}(kr_{\rm fin, ref})\right]^2,
\label{eq:delzfin}
\end{eqnarray}
where $P_{\rm den}(k)$ is the density power spectrum, and $j_{\rm B}^{(1)}$ is the first spherical Bessel function. The derivation of Equation \ref{eq:delzfin} is given in Appendix \ref{App}. In this paper, we use $r_{\rm fin, ref}=10^{-4}$ pc as the same as that of simulations.
To calculate $\left< |\delta v_{\rm ini}|^2 \right>$, we first identify a spherical shell in the final state and then trace the particles belonging to that shell back to the initial state. As shown in Figure \ref{fig:lag}, these shells no longer have spherical symmetry at the initial state and are elongated along the longitudinal axis of the filament (Figure \ref{fig:rate} (b)). Here, for simplicity, we assume that the initial shape of the shells is a cylinder with a radius $\xi_{\rm ini}$ of the cylindrical coordinate and a length of $L_{\rm ini}$. In this case, we can easily calculate $\left< |\delta v_{\rm ini}|^2 \right>$, and it is given as
\begin{eqnarray}
\left< |\delta v_{z, {\rm ini}}|^2 \right> \simeq \frac{1}{3}\sum_{\bm k} P_{\rm vel}(k)\left[  \left\{ F({\bm k}, r_{\rm fin}) - F({\bm k}, r_{\rm fin, ref}) \right\}^2 + F({\bm k}, r_{\rm fin})  F({\bm k}, r_{\rm fin, ref}) k_z^2\left< |\delta z_{\rm ini}|^2 \right> \right].
\label{eq:delv}
\end{eqnarray}
The detailed derivations of Equation \ref{eq:delv} and description of $F({\bm k}, r_{\rm fin})$ are given in Appendix \ref{App}. Using Equations \ref{eq:ens}, \ref{eq:delzfin}, and \ref{eq:delv}, we can derive the following equation:
\begin{eqnarray}
\left< |\delta z_{\rm ini}|^2 \right> \simeq \frac{\left< |\delta z_{\rm fin}|^2 \right>+V_0^2 t^2}{1-W_0^2t^2},
\label{eq:delzini}
\end{eqnarray}
where
\begin{eqnarray}
V_0^2 = \sum_{\bm k} \frac{P_{\rm vel}(k)}{3} \left\{ F({\bm k}, r_{\rm fin}) - F({\bm k}, r_{\rm fin,ref})\right\}^2,
\label{eq:v0}
\end{eqnarray}
\begin{eqnarray}
W_0^2 = \sum_{\bm k} \frac{P_{\rm vel}(k)}{3}  F({\bm k}, r_{\rm fin})F({\bm k}, r_{\rm fin,ref}))k_z^2.
\label{eq:w0}
\end{eqnarray}
From Equation \ref{eq:delzini}, we can estimate the initial difference in the center-of-mass, $\Delta z_{\rm ini} \equiv \sqrt{\left< |\delta z_{\rm ini}|^2 \right>}$, and compare with the results from the simulations. We use $t=5.5 t_{\rm ff}$ measured in our simulations. We need $P_{\rm den}(k)$ and $P_{\rm vel}(k)$ to calculate $\left< |\delta z_{\rm fin}|^2 \right>$ and $\left< |\delta v_{z, {\rm ini}}|^2 \right>$, respectively. Recent observations show that the power spectrum of velocity fluctuations along the filament crest is compatible with the Kolmogorov power spectrum \citep{Arzoumanian2022} and that the velocity dispersion in the low-mass filament is subsonic or transonic \citep{Hacar2011,Hacar2016,Arzoumanian2013}. Therefore, we adopt $P_{\rm vel}(k)d^3k=A_{\rm vel}k^{-11/3}d^3k$ and $\sigma_{\rm 3D}=2c_{\rm s}$, which is the same with parameters in our simulations. Observations also show that the density power spectrum along the filament longitudinal axis follows a Kolmogorov-like spectrum \citep{Roy2015}. Numerical simulations that investigate the properties of density fluctuations generated by the turbulent velocity field suggest that the density power spectrum is compatible with the Kolmogorov slope if the turbulent velocity field is transonic \citep{Kim2005,Pan2022}. Moreover, \cite{Inutsuka2001} shows that if the density power spectrum has the Kolmogorov slope, the resultant core mass function has the Salpeter-like slope in the high-mass end, which is also compatible with the observation of the core mass function \citep{Konyves2015}. Therefore, in this paper, we adopt $P_{\rm den}(k)d^3k=A_{\rm den}k^{-11/3}d^3k$. The dispersion of the density fluctuations, $\sigma_{\rm den} \equiv \sum_{\bm k} P_{\rm den}(k)$, depends on the Mach number,$\mathscr{M}$, \citep{Federrath2010,Pan2016}:
\begin{eqnarray}
\sigma_{\rm den} = b \mathscr{M},
\label{eq:sigmad}
\end{eqnarray}
where $b$ is the forcing parameter. \cite{Kobayashi2022} performed numerical simulations of supersonic converging flows of the warm neutral medium to investigate the turbulence structure of molecular clouds. They show that the solenoidal mode dominates the turbulent velocity field in the molecular cloud and the resultant forcing parameter $b=0.39$. We also measure $\sigma_{\rm d}$ in our simulations and the resultant forcing parameter $b=0.34$ at $\rho_{c}\simeq 10^{-18} \ {\rm g \ cm^{-3}}$. This value is not far from the molecular cloud formation simulations \citep{Kobayashi2022}. Here, we adopt $b=0.34$, but note that our final result $\left< |\delta z_{\rm ini}|^2 \right>$ does not strongly depend on the choice of $b$, since $\left< |\delta z_{\rm ini}|^2 \right>$ is mainly determined by $\left< |\delta v_{z, {\rm ini}}|^2 \right>$. Figure \ref{fig:ana} compares the model and simulation results. The difference between the analytical model and simulations is almost within a factor of 2, although we use the simple model and make several assumptions. This result indicates that the origin of the resultant isotopic profile could be attributed to the initial difference of the center-of-mass of the shells.

\subsection{Isotopic Heterogeneity in Circumstellar Disks}\label{subsec:disk}

\begin{figure*}
\gridline{\fig{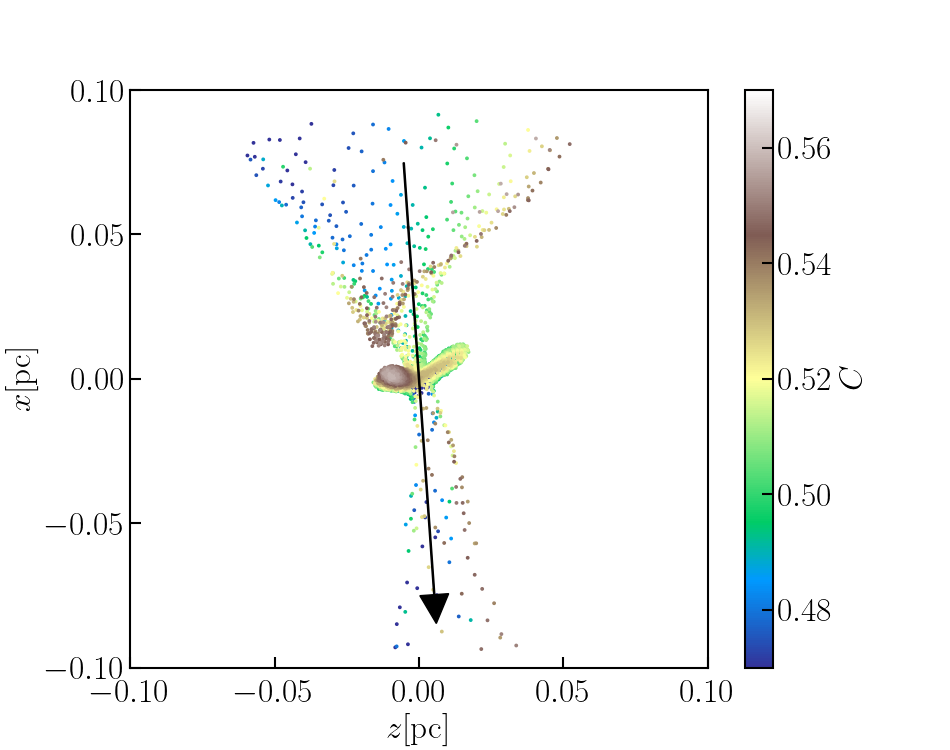}{0.45\textwidth}{(a)}
          \fig{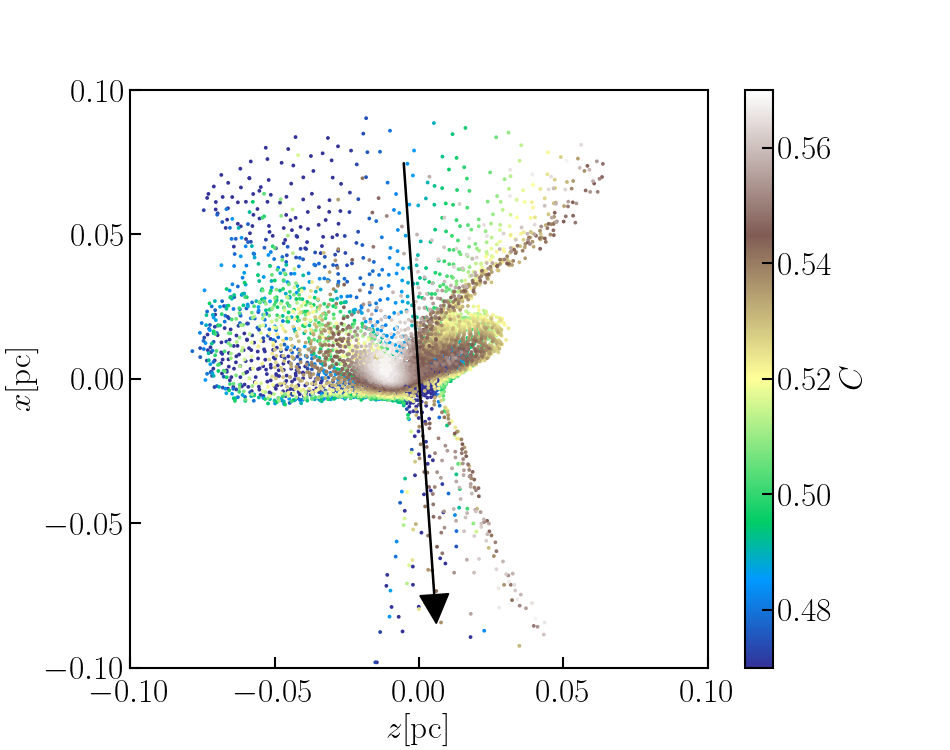}{0.45\textwidth}{(b)}}
\caption{Distribution of the SPH particles with $r_{\rm cf} = 180$-$210$ AU (a) and $r_{\rm cf} = 360$-$400$ AU (b) (Equation \ref{eq:centri}). The color represents the isotopic ratio of the SPH particle. The black arrow represents the rotation direction of the core at a radius of 0.1 pc. \label{fig:rcflag}}
\end{figure*}

\begin{figure}[t]
\plotone{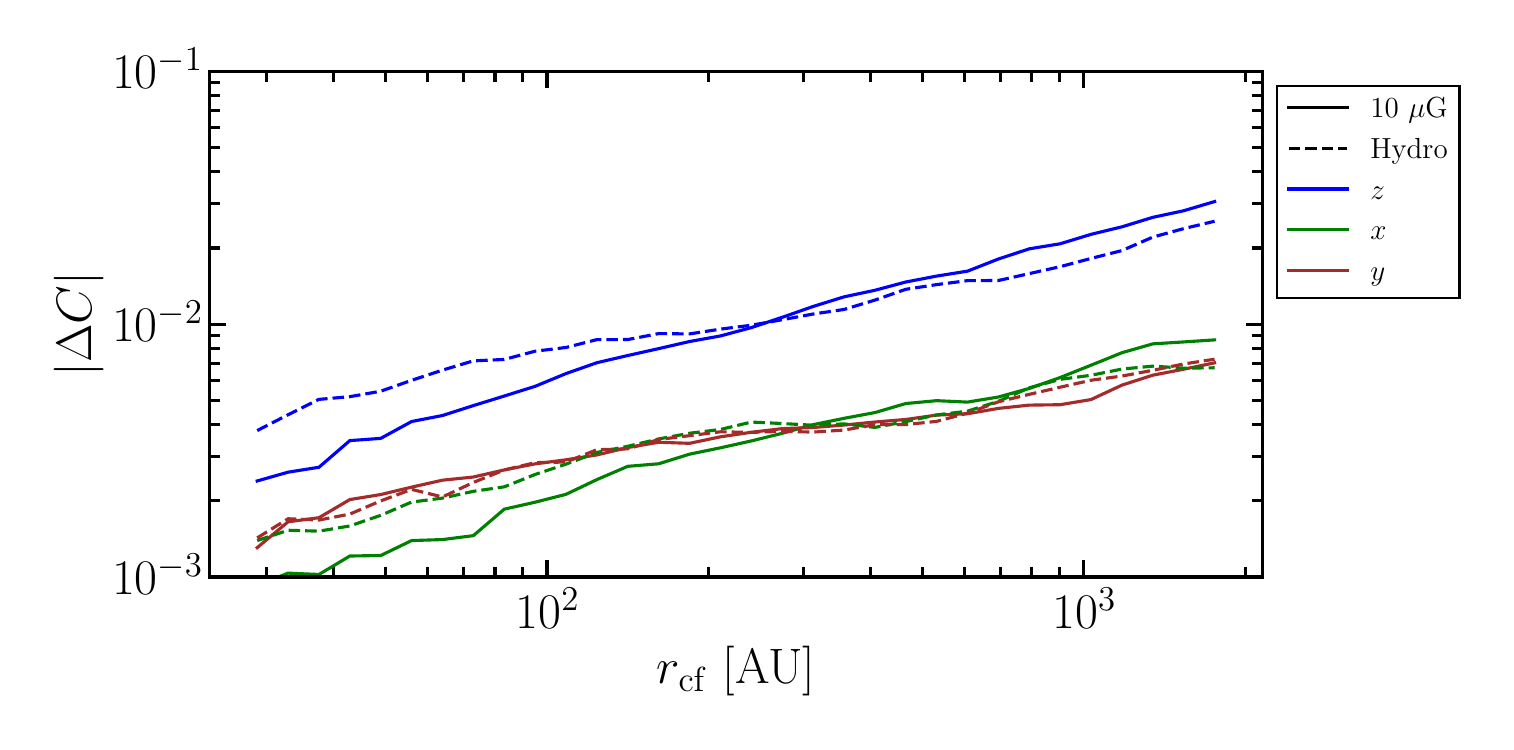}
\caption{Profile of the isotopic ratio in the protoplanetary disks. The horizontal axis is the centrifugal radius estimated from Equation \ref{eq:centri}. The profile is derived using the results at the final state of our simulations. \label{fig:crcf}}
\end{figure}

In section \ref{sec:res}, we discussed the isotopic heterogeneity
 of the molecular cloud cores by assuming the initial isotopic profile in the filament. However, since the planets are thought to form in the circumstellar disk, the isotopic heterogeneity of the circumstellar disk should be investigated. Although our simulations are terminated just before the first core formation, in this subsection, we investigate the distribution of the isotopic ratio in the disk using a simple model. We estimate the radius onto which the fluid elements (SPH particles) fall using the centrifugal radius:
\begin{eqnarray}
r_{{\rm cf},i} = \frac{j_i^2}{GM_{\rm enc}},
\label{eq:centri}
\end{eqnarray}
where $i$ is a label of SPH particle, and $j_{i}$ is the specific angular momentum of the $i$-th SPH particle. $M_{\rm enc}$ is the enclosed mass within the spherical shell with the radius $r_i$, where $r_i$ is the spherical coordinate of the $i$-th SPH particle. The resultant accretion onto the disk is anisotropic since the distribution of the specific angular momentum in the core is anisotropic due to the turbulent velocity field. The spatial distribution of SPH particles that fall onto the disk is illustrated in Figure \ref{fig:rcflag}. Figure \ref{fig:rcflag} (a) displays the SPH particles with the centrifugal radius $r_{\rm cf} = 180$-$210$ AU. In the outer region, the SPH particles are mainly located along the polar direction since they have a smaller angular momentum. The distribution of the SPH particles is relatively axisymmetric around the rotation axis. We also plot the SPH particles with the centrifugal radius $r_{\rm cf} = 360$-$400$ AU in Figure \ref{fig:rcflag} (b). As shown in Figure \ref{fig:rcflag}, even in the thin ring region in the disk, the gas is accreted from various spherical shells of the core. The other disk formation model is considered in Appendix \ref{App_shell} \par
Figure \ref{fig:crcf} shows the profile of the isotopic ratio derived using Equation \ref{eq:centri}. Figure \ref{fig:crcf} indicates that the isotopic ratio remains 0.1 - 1\% of the entire filament, even if we take the effect of the angular momentum distribution in the cores into account. The profile does not show the dependence on the strength of the magnetic field. On the other hand, the profile depends on whether the initial isotopic ratio variation is along the longitudinal or minor axes of the filament. This feature is inherited from the profile in the core, as shown in Figure \ref{fig:Cprofile_model}.

\subsection{Isotopic Heterogeneity in Shells}\label{subsec:disp}

\begin{figure}[t]
\plotone{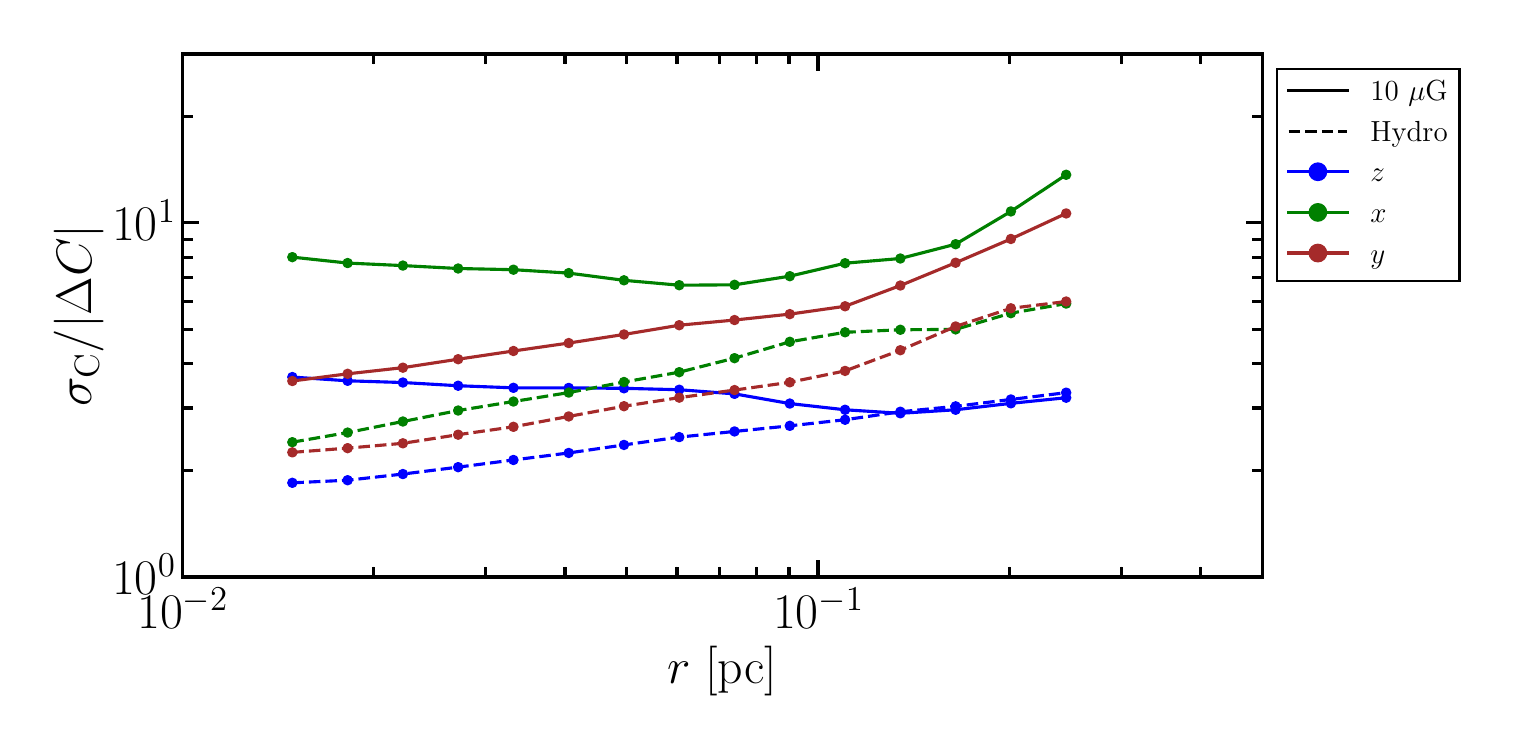}
\caption{Ratio of the dispersion of the isotopic ratio (Equation \ref{eq:dispinshell}) to the average difference of isotopic ratio between shells. \label{fig:deltac}}
\end{figure}

So far, we have discussed the isotopic heterogeneity between the shells (Section \ref{sec:res} and \ref{subsec:ana}) and in the circumstellar disk (Section \ref{subsec:disk}). This means that the inhomogeneities estimated in previous sections are averaged in the polar and azimuthal directions (Equation \ref{eq:shell}). However, recent observations show that the asymmetric structure exists around a protostar and is often interpreted as accretion to the protostar and disk \citep[e.g., ][]{Pineda2020}. This kind of accretion can be reproduced in numerical simulations \citep[e.g., ][]{Kuffmeier2019}. Recently, \cite{Takaishi2024} showed that an unipolar outflow is launched in the magnetized turbulent molecular cloud cores using numerical simulations. Since the unipolar outflow ejects the gas on one side of the core, accretion would no longer be isotropic. Therefore, it is worth estimating the isotopic inhomogeneity in the shell. To do so, we calculate the variation in the shell as follows:
\begin{eqnarray}
\sigma_{C}^2(r)=\frac{\int\int [C(r, \theta, \varphi)-\bar{C}(r)]^2 \rho(r, \theta, \varphi) \sin \theta d \theta d \varphi}{\int\int \rho(r, \theta, \varphi) \sin \theta d \theta d \varphi}.
\label{eq:dispinshell}
\end{eqnarray}
The ratio of $\sigma_{C} (r)$ to $|\Delta C (r)|$ is plotted in Figure \ref{fig:deltac}. Figure \ref{fig:deltac} shows that the dispersion of the isotopic ratio is larger than the average difference in the isotopic ratio between shells, irrespective of the model. The ratio of the model M\_z is smaller than that of models M\_x and M\_y, especially in the outer region. This is because $|\Delta C|$ of models M\_x and M\_y is smaller than that of model M\_z. $\sigma_C$ itself is a monotonically increasing function since $\sigma_C$ comes from the initial linear gradient of the isotopic ratio that we assumed. This result indicates that the anisotropic infall would make an impact on the isotopic inhomogeneity in the circumstellar disk. 
\section{Summary \& Conclusion} \label{sec:sum}
In this paper, we investigate the isotopic heterogeneity of molecular cloud cores formed through filament fragmentation using numerical simulations. Our results are summarized as follows.
\begin{enumerate}
\item In our model, the isotopic heterogeneity within the filament is normalized to a scale from 0 to 1, and we calculate the resulting heterogeneity that is brought into the core. Specifically, we divide the core into concentric shells and estimate the difference in isotopic ratios between each shell and the core center. As a result, we find that the core retains approximately 0.1–1\% of the initial heterogeneity present over the filament length of 1 pc.
\item We investigate cases where isotopic heterogeneity exists along both the major and minor axes of the filament. The results show that an isotopic ratio variation along the major axis leads to greater heterogeneity in the core. This is because the core accretes more gas from the direction of the major axis.
\item In the final state, each shell shows a different isotopic ratio relative to the core center. We find that this is due to variations in the initial center-of-mass positions of the shells caused by the turbulence.
\item Although our simulation stops just before the formation of the first core, we also estimate isotopic heterogeneity in the disk using a model based on the centrifugal radius. The results indicate that while turbulence induces anisotropy in angular momentum, about 1\% of the initial isotopic heterogeneity over the filament length of 1 pc remains within the disk.
\end{enumerate}

Combining our results with those of \cite{Jacquet2019}, the following picture emerges: about 0.1–1\% of the isotopic anomalies present in the filament are delivered to the dense core, and roughly one-third of that heterogeneity is then transferred from the core to the disk. Therefore, the residual heterogeneity from the filament that remains in the disk is approximately 0.03–0.3\%, which is sufficient to explain the isotopic anomalies observed in ratios such as $^{54}$Cr$/^{52}$Cr and $^{50}$Ti$/^{47}$Ti \citep{Dauphas2016}. As shown in Section \ref{subsec:disp}, the anisotropic would play a role in determining the distribution of the isotope in the disk. A high-resolution simulation is required to investigate the realistic properties of the isotopic ratio in the circumstellar disk, and it is our future work.

\begin{acknowledgments}
Numerical computations were carried out on Cray XC50 at the Center for Computational Astrophysics, National Astronomical Observatory of Japan. This work is supported by JSPS KAKENHI grant No. 23K19073, 18H05436, and 18H05437.
\end{acknowledgments}

\software{Matplotlib \citep{Hunter2007}, Numpy \citep{Harris2020} }

\appendix

\section{Derivations of Correlation Functions}\label{App}
In this appendix, we derive Equations \ref{eq:delzfin} and \ref{eq:delv}. First, $\left< |\delta z_{\rm fin}|^2 \right>$ is defined as follows: 
\begin{eqnarray}
\left< |\delta z_{\rm fin}|^2 \right> = \left< |\bar{z}_{\rm c}(r_{\rm fin}) - \bar{z}_{\rm c}(r_{\rm fin,ref}) |^2 \right>,
\label{eq:delzfindef}
\end{eqnarray}
where $\left<A\right>$ and $\bar{A}$ denote the ensemble average and the average in the shell as written in Equation \ref{eq:shell}, respectively. First, we estimate $\bar{z}_{\rm c}$ from the following expression:
\begin{eqnarray}
\bar{z}_{\rm c}(r_{\rm fin}) = \frac{1}{\rho_0V_{\rm fin}}\int z(\rho_0+\delta \rho)dV.
\label{eq:barzfindef}
\end{eqnarray}
$\delta \rho$ represents the density fluctuations induced by the initial Kolmogorov turbulence. For simplicity, we assume constant density in the core, $\rho_0$. $V_{\rm fin}$ denotes the volume of the shell, $V_{\rm fin}=4\pi r_{\rm fin}^2\delta r$. $\delta r$ is the thickness of the shell. The integration on the right-hand side of Equation \ref{eq:barzfindef} can be done in the polar and azimuthal directions and can be written as follows:
\begin{eqnarray}
\bar{z}_{\rm c}(r_{\rm fin})  = ir_{\rm fin} \sum_{\bm k} \frac{\widetilde{\delta \rho}_{\bm k}}{\rho_0}j_{\rm B}^{(1)}(kr_{\rm fin}),
\label{eq:barzfin}
\end{eqnarray}
where $\widetilde{A}_{\bm k}$ denotes the Fourier component of $A$,
\begin{eqnarray}
A({\bm x}) = \sum_{\bm k} \widetilde{A}_{\bm k}\exp(i{\bm k}\cdot{\bm x}).
\label{eq:ft}
\end{eqnarray}
$j_{\rm B}^{(1)}(X)$ is the spherical Bessel function defined as \citep{Abramowitz1965}
\begin{eqnarray}
j_{\rm B}^{(1)}(X) = -\frac{i}{2}\int^{\pi}_{0}\exp(iX\cos\theta)\cos\theta \sin\theta d\theta.
\label{eq:sb1}
\end{eqnarray}
We also define the density power spectrum as follows:
\begin{eqnarray}
P_{\rm den}(k) = \left< \left| \frac{\widetilde{\delta \rho}_{\bm k}}{\rho_0}\right|^2 \right>.
\label{eq:dp}
\end{eqnarray}

Using Equations \ref{eq:delzfindef}, \ref{eq:barzfin}, and \ref{eq:dp}, we can derive Equations \ref{eq:delzfin}. \par
Equations \ref{eq:delv} can also be derived in a similar manner:
\begin{eqnarray}
\left< |\delta v_{z, {\rm ini}}|^2 \right> = \left< |\bar{v}_{z,{\rm c}}(\xi_{\rm ini},L_{\rm ini},\delta z_{\rm ini}) - \bar{v}_{z,{\rm c}}(\xi_{\rm ini,ref},L_{\rm ini,ref},\delta z_{\rm ini,ref}) |^2 \right>,
\label{eq:delvzinidef}
\end{eqnarray}
where $\xi_{\rm ini}$ and $L_{\rm ini}$ are the radius of the cylindrical coordinate and the length of the cylinder. Since $|\delta z_{\rm ini}| \gg |\delta x_{\rm ini}|, |\delta y_{\rm ini}| $ as shown in Figure \ref{fig:center}, here we consider only $|\delta z_{\rm ini}|$. In addition, we assume that the length of the cylinder can be described as $L_{\rm ini} = f_{\rm ar}\xi_{\rm ini}$. Based on the linear analysis for the filament fragmentation \citep{Inutsuka1992} and simulation results (Figure \ref{fig:rate} (b)), we adopt $f_{\rm ar}=4.0$. Since $\delta z_{\rm ini}$ is defined as the difference of the center-of-mass with respect to the reference shell, now we can set $\delta z_{\rm ini,ref}=0$. Equation \ref{eq:delvzinidef} is simplified as follows:
\begin{eqnarray}
\left< |\delta v_{z, {\rm ini}}|^2 \right> = \left< |\bar{v}_{z,{\rm c}}(\xi_{\rm ini},\delta z_{\rm ini}) - \bar{v}_{z,{\rm c}}(\xi_{\rm ini,ref}) |^2 \right>.
\label{eq:delvzinidef2}
\end{eqnarray}
$\bar{v}_{z,{\rm c}}(\xi_{\rm ini},\delta z_{\rm ini})$ is estimated from the following equation:
\begin{eqnarray}
\bar{v}_{z,{\rm c}}(\xi_{\rm ini},\delta z_{\rm ini}) = \frac{1}{\rho_0V_{\rm ini} }\int \rho_0 v_z({\bm x} + \delta z_{\rm ini} {\bm e}_z)dV,
\label{eq:vzcini}
\end{eqnarray}
where $V_{\rm ini}$ denotes the shell of cylinder, $V_{\rm ini}=3\pi f_{\rm ar} \xi^2_{\rm ini} \delta \xi_{\rm ini}= 12\pi \xi_{\rm ini}^2 \delta \xi_{\rm ini}$. Using Equation \ref{eq:ft}, Equation \ref{eq:vzcini} can be written as
\begin{eqnarray}
\bar{v}_{z,{\rm c}}(\xi_{\rm ini},\delta z_{\rm ini}) = \frac{2}{3} \sum_{\bm k} \widetilde{v}_{z, {\bm k}} \exp(ik_z \delta z_{\rm ini}) \left[J_{\rm B}^{(0)}(k_{\xi}\xi_{\rm ini})j_{\rm B}^{(0)}\left( \frac{k_zL_{\rm ini}}{2}\right) - \frac{J_{\rm B}^{(1)}(k_{\xi}\xi_{\rm ini})}{k_{\xi}\xi_{\rm ini}} \cos\left( \frac{k_zL_{\rm ini}}{2}\right)\right],
\label{eq:vzcini2}
\end{eqnarray}
where we have used the following equations \citep{Abramowitz1965}:
\begin{eqnarray}
j_{\rm B}^{(0)}\left(\frac{k_zL_{\rm ini}}{2}\right) = \frac{1}{L_{\rm ini}}\int^{+L_{\rm ini}/2}_{-L_{\rm ini}/2}\exp(ik_zz) dz,
\label{eq:sb0}
\end{eqnarray}

\begin{eqnarray}
J_{\rm B}^{(0)}\left({k_{\xi}\xi}\right) = \frac{1}{2\pi}\int^{+\pi}_{-\pi}\exp(ik_\xi \xi \cos \phi) d\phi,
\label{eq:b0}
\end{eqnarray}

\begin{eqnarray}
J_{\rm B}^{(1)}\left(k_{\xi}\xi\right) = \frac{k_{\xi}}{\xi_{\rm ini}}\int^{\xi_{\rm ini}}_{0} J_{\rm B}^{(0)}\left({k_{\xi}\xi}\right) \xi d\xi.
\label{eq:b1}
\end{eqnarray}
To calculate $\bar{v}_{z,{\rm c}}(\xi_{\rm ini},\delta z_{\rm ini})$ as a function of $r_{\rm fin}$, we need to know the relation between $\xi_{\rm ini}$ and $r_{\rm fin}$, $\xi_{\rm ini}=\mathscr{R}(r_{\rm fin})$. The detailed description of $\mathscr{R}(r_{\rm fin})$ is given in Appendix \ref{Appr}. Once $\mathscr{R}(r_{\rm fin})$ is given, we define $F({\bm k}, r_{\rm fin})$ as follows:
\begin{eqnarray}
F({\bm k}, r_{\rm fin}) \equiv \frac{2}{3} \left[J_{\rm B}^{(0)}(k_{\xi}\xi_{\rm ini})j_{\rm B}^{(0)}\left( \frac{k_zL_{\rm ini}}{2}\right) - \frac{J_{\rm B}^{(1)}(k_{\xi}\xi_{\rm ini})}{k_{\xi}\xi_{\rm ini}} \cos\left( \frac{k_zL_{\rm ini}}{2}\right)\right]
\label{eq:fkr}
\end{eqnarray}
Using this expression, the velocity of shells is rewritten as
\begin{eqnarray}
\bar{v}_{z,{\rm c}}(r_{\rm fin},\delta z_{\rm ini}) =\sum_{\bm k} \widetilde{v}_{z, {\bm k}} \exp(ik_z \delta z_{\rm ini}) F({\bm k}, r_{\rm fin}),
\label{eq:vzcinif}
\end{eqnarray}
\begin{eqnarray}
\bar{v}_{z,{\rm c}}(r_{\rm fin,ref}) =\sum_{\bm k} \widetilde{v}_{z, {\bm k}} F({\bm k}, r_{\rm fin,ref}).
\label{eq:vzcinireff}
\end{eqnarray}
Note that $k_z \delta z_{\rm ini} \ll 1$ holds in the longer wavelength that has a large portion of the power of the turbulent velocity field. In Equation \ref{eq:vzcinif}, these longer wavelengths largely contribute to the resultant $\bar{v}_{z,{\rm c}}(r_{\rm fin},\delta z_{\rm ini})$. Therefore, Equation \ref{eq:vzcinif} can be approximated as
\begin{eqnarray}
\bar{v}_{z,{\rm c}}(r_{\rm fin},\delta z_{\rm ini}) \simeq \sum_{\bm k} \widetilde{v}_{z, {\bm k}} (1 + ik_z \delta z_{\rm ini}) F({\bm k}, r_{\rm fin}),
\label{eq:vzcinifex}
\end{eqnarray}

By substituting Equations \ref{eq:vzcinifex} and \ref{eq:vzcinireff} into Equation \ref{eq:delvzinidef2}, we can derive Equation \ref{eq:delv}. We have used $\left< \left| \delta v_{z,{\bm k}}\right|^2 \right> = P_{\rm vel}(k)/ 3$ because of the isotropy of the Kolmogorov turbulent field and, for simplicity, we ignore the correlation between $\widetilde{v}_{z, {\bm k}}$ and $\delta z_{\rm ini}$.

\setcounter{equation}{0}
\section{Derivations of Size Relation}\label{Appr}
To derive $\mathscr{R}(r_{\rm fin})$, we consider mass conservation. At the initial state, since we now consider the cylindrical region, the contained mass in that region can be written as
\begin{eqnarray}
M=0.47 f_{\rm ar} M_{\odot}\frac{{\xi '}_{\rm ini}^{3}}{1+{\xi '}_{\rm ini}^{2}},
\label{eq:massini}
\end{eqnarray}
where ${\xi '}_{\rm ini}=\xi_{\rm ini}/H_0$, $H_0$ is the scale height of the filament as shown in Equation \ref{eq:scaleheight}. On the other hand, the density profile within the core at the final state is discussed in Section 3.2 of \cite{Misugi2024}, the region focused in this paper roughly corresponds to the intermediate region between the self-similar region ($\lesssim 1000$ AU) and filament region ($\gtrsim 0.1$ pc). The self-similar solution known as Larson-Penston solution \citep{Larson1969,Penston1969} is
\begin{eqnarray}
M_{\rm lp}=0.3978 M_{\odot} \left( \frac{r_{\rm fin}}{1000\ {\rm AU}}\right) .
\label{eq:masslp}
\end{eqnarray}
On the other hand, at the filament scale, the filament can be regarded as the one-dimensional structure, and it is given as
\begin{eqnarray}
M_{\rm f} = 3.72\ M_{\odot}\left( \frac{r_{\rm fin}}{0.1\ {\rm pc}}\right).
\label{eq:massfil}
\end{eqnarray}
By interpolating between these two regions, we can get the following expression:
\begin{eqnarray}
M = 3.72\ M_{\odot}\left( \frac{r_{\rm fin}}{0.1\ {\rm pc}}\right)^{0.74}.
\label{eq:massintp}
\end{eqnarray}
Using Equations \ref{eq:massini}, \ref{eq:massintp} and the mass conservation, we can derive the following equation:
\begin{eqnarray}
\frac{{\xi '}_{\rm ini}^{3}}{1+{\xi '}_{\rm ini}^{2}} = \frac{4.8}{f_{ar}} {r'}_{\rm fin}^{0.74}.
\label{eq:massinp}
\end{eqnarray}
The solution of Equation \ref{eq:massinp} is given as
\begin{eqnarray}
{\xi '}_{\rm ini} = -\frac{p^2}{3\sqrt[3]{q}} + \frac{p}{3} - \frac{\sqrt[3]{q}}{3},
\label{eq:massinpsol}
\end{eqnarray}
where
\begin{eqnarray}
p \equiv \frac{4.8}{f_{ar}} {r'}_{\rm fin}^{0.74},
\label{eq:massinpa}
\end{eqnarray}
\begin{eqnarray}
q \equiv -p^3 - \frac{27p}{2} + \frac{p}{2}\sqrt{-4p^4 + (2p^2 + 27)^2}.
\label{eq:massinpb}
\end{eqnarray}
Equation \ref{eq:massinp} can be formally written as $\xi_{\rm ini}=\mathscr{R}(r_{\rm fin})$.

\section{Shell Model}\label{App_shell}
\begin{figure*}

\gridline{\fig{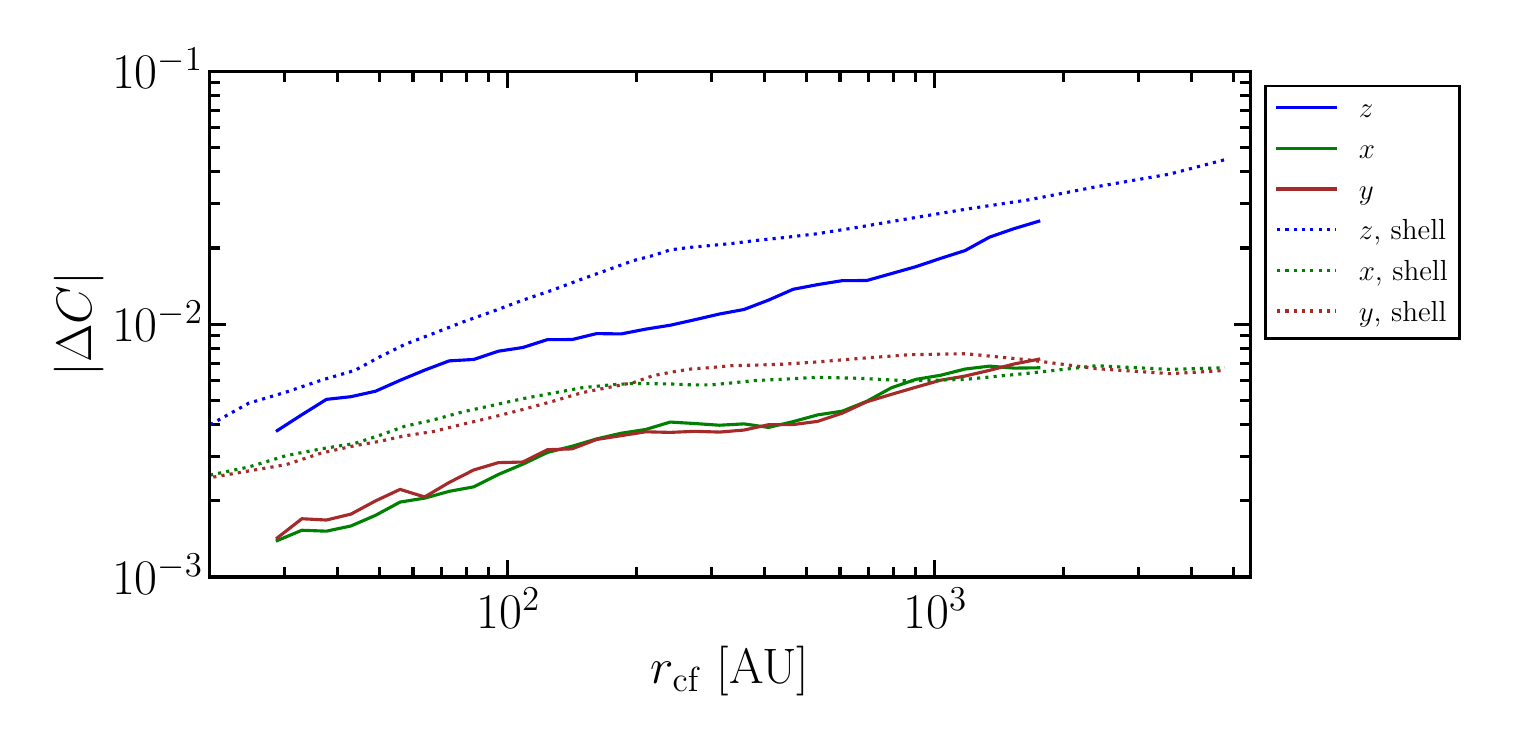}{0.48\textwidth}{(a)}
          \fig{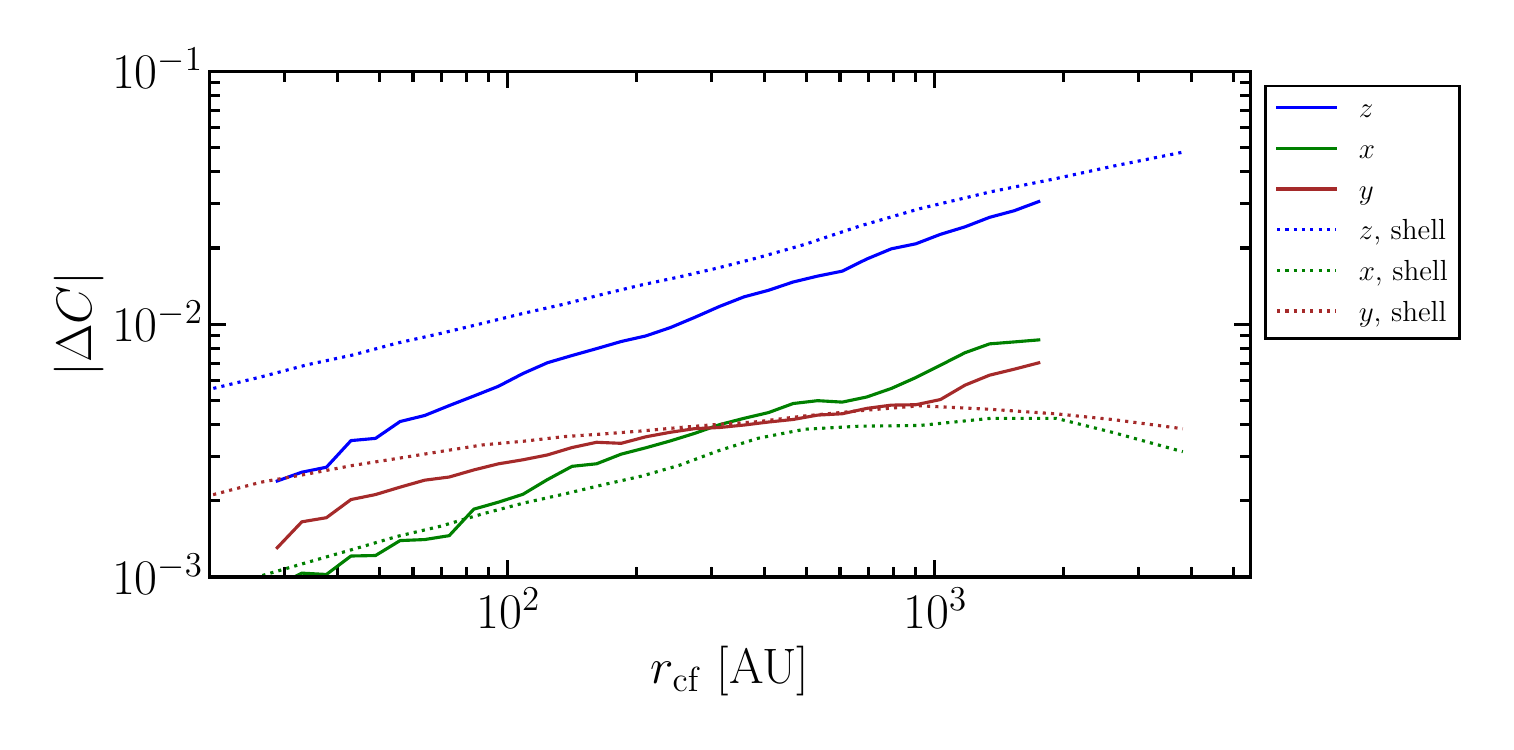}{0.48\textwidth}{(b)}}

\caption{Profile of the isotopic ratio in the protoplanetary disks. The horizontal axis is the centrifugal radius estimated from
Equation \ref{eq:centrsh} . The profile is derived using the results at the final state of our simulations in (a) hydrodynamic and (b) $B_0=10{\rm \mu G}$ cases
\label{fig:shell}}
\end{figure*}

In this section, we present the results of isotopic inhomogeneity in the circumstellar disk obtained using the other extreme model. In the model adopted in Section \ref{subsec:disk}, Equation \ref{eq:centri} is used, such that particles fall onto the circumstellar disk according to the angular momentum of each fluid element in the shell. As a result, the isotopic ratio in the disk is determined by a combination of particles originating from different spherical shells. We examine how the results presented in Section \ref{subsec:disk} depend on the choice of disk formation model.
Here, we consider altanative model by assuming that each spherical shell falls onto the disk at a radius determined by its angular momentum, given by
\begin{eqnarray}
r_{{\rm cf,shell}} = \frac{j_{\rm shell}^2}{GM_{\rm enc}},
\label{eq:centrsh}
\end{eqnarray}
where $j_{\rm shell}$ is the specific angular momentum of a spherical shell measured at the final state of our simulations. Figure \ref{fig:shell} compares the results obtained from the two models. The isotopic inhomogeneity resulting from Equation \ref{eq:centrsh} is slightly larger than that obtained from Equation \ref{eq:centri}. This difference arises because, from Equation \ref{eq:centrsh}, a given spherical shell falls onto a single, specific disk radius while particles originating from a given spherical shell fall onto a broad range of disk radii when Equation \ref{eq:centri} is adopted. However, this difference is only factor of 2 or 3. This result indicates that the radial profile of isotopic inhomogeneity in the circumstellar disk does not strongly depend on the choice of the disk formation model.

\bibliography{sample631}{}
\bibliographystyle{aasjournal}

\end{document}